\documentclass[reqno,draft]{amsart}
\usepackage{amssymb}
\usepackage{upref}
\newcommand{\bbN}{{\mathbb{N}}}
\newcommand{\bbR}{{\mathbb{R}}}

\newcommand{\bbC}{{\mathbb{C}}}
\newcommand{\bbCinf}{{\mathbb{C}_{\infty}}}

\newcommand{\calF}{{\mathcal F}}

\newcommand{\calI}{{\mathcal I}}
\newcommand{\calJ}{{\mathcal J}}
\newcommand{\calK}{{\mathcal K}}

\newcommand{\no}{\nonumber}
\newcommand{\lb}{\label}
\newcommand{\bi}{\bibitem}
\newcommand{\f}{\frac}
\newcommand{\ul}{\underline}
\newcommand{\ol}{\overline}
\newcommand{\dott}{\,\cdot\,}
\newcommand{\ti}{\tilde} % use only for capital letters
\newcommand{\wti}{\widetilde  }
\newcommand{\uF}{\breve{F}}
\newcommand{\uP}{\breve{P}}
\newcommand{\uR}{\breve{R}}
\newcommand{\uK}{\breve{{\mathcal K}}}

\renewcommand{\Re}{\text{\rm Re}}
\renewcommand{\Im}{\text{\rm Im}}

\DeclareMathOperator{\KdV}{KdV}
\DeclareMathOperator{\sKdV}{s-KdV}

\allowdisplaybreaks
\numberwithin{equation}{section}

\newtheorem{theorem}{Theorem}[section]
\newtheorem{lemma}[theorem]{Lemma}

\newtheorem{hypothesis}[theorem]{Hypothesis}
\theoremstyle{definition}

\theoremstyle{remark}
\newtheorem{remark}[theorem]{Remark}

\begin{document}

\title[Infinite genus limit]{Integrable Systems in the
infinite genus limit}

% Information for first author
\author[Gesztesy]{Fritz Gesztesy}
\address{Department of Mathematics,
University of Missouri,
Columbia, MO 65211, USA}
\email{fritz@math.missouri.edu}
\urladdr{http://www.math.missouri.edu/people/fgesztesy.html}
% Information for second author

%%%%%%%%%%%%%%%%%%%%%%%%%%%%%%%%%%%%%%%%%%%%%%%%%%%%%%%%%%
\dedicatory{Dedicated with great pleasure to Sergio
Albeverio on the occasion of his 60th birthday}
%%%%%%%%%%%%%%%%%%%%%%%%%%%%%%%%%%%%%%%%%%%%%%%%%%%%%%%%%%

\thanks{Supported in part by the
University of Missouri Research Board grant RB-97-086.}
\date{\today}
%\keywords{KdV hierarchy, Darboux transformations,
%hyperelliptic curves}
%\subjclass{Primary 35Q53, 35Q55, 58F07; Secondary 35Q51, 35Q58}

%%%%%%%%%%%%%%%%%%%%%%%%%%%%%%%%%%%%%%%%%%%%%%%%%%%%%%%%%%%%%%%%%%%
\begin{abstract}
We provide an elementary approach to integrable systems associated
with hyperelliptic curves of infinite genus. In particular, we
explore the extent to which the classical Burchnall-Chaundy theory
generalizes in the infinite genus limit, and systematically study
the effect of Darboux transformations for the KdV hierarchy on such
infinite genus curves. Our approach applies to complex-valued
periodic solutions of the KdV hierarchy and naturally identifies
the Riemann surface familiar from standard Floquet theoretic
considerations with a limit of Burchnall-Chaundy curves.
\end{abstract}
%%%%%%%%%%%%%%%%%%%%%%%%%%%%%%%%%%%%%%%%%%%%%%%%%%%%%%%%%%%%%%%%%%%

\maketitle

%%%%%%%%%%%%%%%%%%%%%%%%%%%%%%%%%%%%%%%%%%%%%%%%%%%%%%%%%%%%%%%%%%%
\section{Introduction} \lb{s1}
%%%%%%%%%%%%%%%%%%%%%%%%%%%%%%%%%%%%%%%%%%%%%%%%%%%%%%%%%%%%%%%%%%%

Ever since Marchenko's 1974 treatment of the periodic KdV problem
(cf.~\cite{Ma86} and the references therein) and especially after
the work by McKean and Trubowitz in 1976 and 1978 \cite{MT76},
\cite{MT78}, it seemed natural to consider infinite genus limits of
completely integrable systems. In fact, in a series of papers from
1982--1985, Levitan \cite{Le82}, \cite{Le83}, \cite{Le84},
\cite{Le85} addressed the Jacobi inversion problem on infinite
genus hyperelliptic curves and constructed classes of
quasi periodic and almost periodic potentials. The corresponding
constructions of KdV flows can be found in Ch.~12 of his monograph
\cite{Le87}. These investigations of almost periodic Schr\"odinger
operators were continued by Kotani and Krishna \cite{KK88}, Craig
\cite{Cr89}, and many others (see, e.g., the recent work by
Sodin and Yuditskii \cite{SY95} and the references therein). In the
past five years a resurgence of interest in these problems appears
to have taken place, as demonstrated by papers of Egorova
\cite{Eg94}, Feldman, Kn\"orrer, and Trubowitz \cite{FKT96},
\cite{FKT97}, M\"uller, Schmidt, and Schrader \cite{MSS96},
\cite{MSS98}, Schmidt \cite{Sc96}, Zakharevich \cite{Za97}, and
Merkl \cite{Me97}, \cite{Me99}. The various approaches in these
papers, however, are quite different. Some focus on the
construction of theta functions associated with infinite period
matrices for the underlying infinite genus curve (cf.~\cite{FKT96},
\cite{FKT97}, \cite{MSS96}, \cite{MSS98}) and then derive KdV, KP,
etc., solutions in analogy to the Its-Matveev formula \cite{IM75}
in the finite genus case, whereas other authors focus on divisors
corresponding to potentials (cf. \cite{Me97}, \cite{Me99},
\cite{Sc96}, \cite{Za97}) and hence devote their attention to
extensions of the Riemann-Roch theorem to infinite genus curves. In
either approach, the traditional use of Lax pairs to define the
underlying infinite genus Riemann surface is avoided as it formally
enforces the use of differential expressions of infinite order.
Indeed, stationary algebro-geometric KdV solutions $V$ by
definition correspond to a commuting Lax pair $(P_{2n+1},L),$
$[P_{2n+1},L]=0,$ where $L=-\frac{d^2}{dx^2}+V(x)$ is a
second-order Schr\"odinger differential expression with potential
$V(x)$ and $P_{2n+1}$ is a differential expression of order
$2n+1,$ $n\in\bbN\cup\{0\}$ (cf.~Section~\ref{s2} for details).
The Burchnall-Chaundy polynomial associated with
$(P_{2n+1},L)$ then is of the form $P_{2n+1}^2+
R_{2n+1}(L)=0$ for some polynomial $R_{2n+1} (z) =
\prod_{m=0}^{2n} (z-E_m),$ $\{ E_m\}_{m=0,\dots,2n} \subset \bbC,$
of degree $2n+1$ in $z,$ naturally defining the hyperelliptic curve
$\calK_n$ of (arithmetic) genus $n,$
\begin{equation}
\calK_{n}\colon y^2 = \prod_{m=0}^{2n} (z-E_m). \label{1.1}
\end{equation}
Thus, infinite genus KdV potentials $V$ formally correspond to
infinite-order differential expressions $P_\infty,$ not too popular
a subject! However, a moment of reflection reveals that the
situation at hand can be controlled as follows. In the finite genus
case, $P_{2n+1}$ restricted to the two-dimensional nullspace
$\ker(L-z),$ $z\in\bbC$ of $L,$ just becomes a first-order
differential expression of the type,
\begin{equation}
P_{2n+1}\big|_{\ker(L-z)} =
\big(F_n (z,x) \f{d}{dx} -(1/2) F_{n,x}(z,x)\big)
\big|_{\ker(L-z)}, \lb{1.2}
\end{equation}
where $F_n(z,x)$ is a recursively defined polynomial of degree
$n$ with respect to $z\in\bbC$ (cf.~Section~\ref{s2} for details).
The corresponding stationary
$n$th KdV equation satisfied by $V$ is then of the type
\begin{equation}
\sKdV_n(V)=(1/2)F_{n,xxx}(0,x)-2V(x)F_{n,x}(0,x)-V_x(x)F_n(0,x)=0.
\lb{1.3}
\end{equation}
Hence to describe infinite genus situations, where the KdV potential
$V(x)$ corresponds to an infinite genus curve $\calK_\infty$ of
the type
\begin{equation}
\calK_\infty: y^2= (E_0-z) \prod_{m\in\bbN} (1-(z/E_m)),
\quad \{E_m\}_{m\in\bbN}\subset\bbC, \,\,
\sum_{m\in\bbN}|E_m|^{-1}<\infty, \lb{1.4}
\end{equation}
one simply needs to replace the polynomial $F_n(z,x)$ by an
appropriate entire function $F_\infty (z,x)$ and this simple recipe
is a guiding principle for this paper.

As a result we obtain a canonical stationary $\KdV_\infty$ equation
satisfied by any KdV potential $V$ associated with an infinite
genus curve $\calK_\infty$ in \eqref{1.4}. In complete analogy to
the finite genus case \eqref{1.3}, this equation is of the type
\begin{equation}
\sKdV_\infty(V)=(1/2)F_{\infty,xxx}(0,x)-2V(x)F_{\infty,x}(0,x)
-V_x(x)F_\infty (0,x)=0. \lb{1.5}
\end{equation}

As a concrete application of our formalism we show that any
complex-valued periodic $C^1(\bbR)$ potential $V$ satisfies an
equation of the type \eqref{1.5}. Moreover, these considerations
naturally identify the underlying Riemann surface (generically of
infinite genus), obtained from Floquet theoretic considerations,
with the infinite genus limit of Burchnall-Chaundy curves.

Next, a quick outline of the content of each section.
Section~\ref{s2} reviews a construction of the KdV hierarchy; the
infinite genus limit is discussed in detail in our principal
Section~\ref{s3}; the effect of Darboux transformations on infinite
genus curves of the type \eqref{1.4} is studied in our final
Section~\ref{s4}.

Although this paper focuses only on the case of the KdV
hierarchy,  combining the methods of this paper with the polynomial
recursion approach developed in \cite{BGHT98}, \cite{GH97},
\cite{GH99},
\cite{GR98} immediately extends all results to other completely
integrable hierarchies associated with hyperelliptic curves such as
the Toda lattice, sine-Gordon, classical Boussinesq, and AKNS
(nonlinear Schr\"odinger) hierarchies.
%%%%%%%%%%%%%%%%%%%%%%%%%%%%%%%%%%%%%%%%%%%%%%%%%%%%%%%%%%%%%%%%%%%
\section{The KdV hierarchy} \lb{s2}
%%%%%%%%%%%%%%%%%%%%%%%%%%%%%%%%%%%%%%%%%%%%%%%%%%%%%%%%%%%%%%%%%%%

In this section we briefly recall an explicit construction of the
KdV hierarchy.

We start by introducing a polynomial recursion formalism following
Al'ber \cite{Al79}, \cite{Al81} (see also \cite[Ch.\ 12]{Di91},
\cite{GD75}, \cite{GD79}) presented
in detail in \cite{GRT96}, \cite{GW93} (see also \cite{GH98},
\cite{GH00}, \cite{GW96}).

Suppose $V\colon\bbC\to\bbCinf$ (with
$\bbCinf=\bbC\cup\{\infty\}$) is meromorphic and
consider the Schr\"odinger operator
\begin{equation}
L = - \frac{d^2}{dx^2} + V(x), \quad x\in\bbC. \lb{2.1}
\end{equation}
Introducing $\{f_j(x)\}_{j\in\bbN_0}$, with
$\bbN_0=\bbN\cup\{0\}$,  recursively by
\begin{equation}
f_0 = 1,\quad f_{j,x} = -(1/4) f_{j-1, xxx} +
Vf_{j-1,x} + (1/2) V_x f_{j-1},\quad j\in\bbN,
\lb{2.2}
\end{equation}
one finds explicitly,
\begin{align}
f_0 & =1, \no \\
f_1 & = (1/2) V + c_1,  \no \\
f_2 &= - (1/8) V_{xx} +(3/8) V^2 +
 c_1(1/2) V + c_2,  \lb{2.3} \\
f_3 & = (1/32) V_{xxxx}-(5/16)VV_{xx}
-(5/32)V_x^2+(5/16)V^3 \no \\
&\quad \, +c_2 (1/2) V+c_1(-(1/8) V_{xx}+(3/8)V^2) +c_3,
\text{ etc.,} \no
\end{align}
where $\{c_\ell\}_{\ell\in\bbN}\subset\bbC$ denote integration
constants. The ${f}_{j}$ are well-known to be differential
polynomials in
$V$ (see, e.g., \cite{Ei83}).  Given $V$ and $f_{j}$ one
defines  differential expressions $P_{2n+1}$ of order $2n+1$,
\begin{equation}
P_{2n+1}  = \sum_{j=0}^n \big(f_{n-j}(x)
\f{d}{dx}-(1/2) f_{n-j,x} (x) \big) L^{j},
\quad n\in\bbN_0,
\lb{2.6}
\end{equation}
and one verifies
\begin{equation}
[P_{2n+1},L] = 2f_{n+1,x}, \quad n\in\bbN_0,
\lb{2.7}
\end{equation}
with $[\dott,\dott]$ the commutator. The stationary KdV
hierarchy, denoted by $\sKdV_n(\cdot),$ $n\in\bbN_0,$
is then defined in terms of the stationary  Lax relations
\begin{equation}
-[P_{2n+1}, L] =-2f_{n+1,x}=\sKdV_n(V)=0,\quad n\in\bbN_0.
\lb{2.8}
\end{equation}
Explicitly, one finds for the first few s-KdV equations
\begin{align}
n&=0: \quad \sKdV_0(V)=-V_x =0, \no \\
n&=1: \quad \sKdV_1(V)=(1/4) V_{xxx} -(3/2) VV_x - c_1V_x=0,
\lb{2.9}
\\ n&=2: \quad \sKdV_2(V)=-(1/16) V_{xxxxx} +(5/8) VV_{xxx}
+(5/4) V_x V_{xx} - (15/8) V^2 V_x \no \\
& \hspace*{1.25cm}
 - c_2 V_x + c_1 ((1/4) V_{xxx} -(3/2) VV_x)=0,
\text{  etc.} \no
\end{align}
$V(x)$ is called an {\it algebro-geometric} KdV {\it potential} if
it satisfies one (and hence infinitely many) of the equations in
the  stationary KdV hierarchy \eqref{2.8}.

Introducing the fundamental polynomial $F_{n}(z,x)$ of
degree $n$ in $z\in\bbC,$
\begin{equation}
F_n (z,x)  = \sum_{j=0}^n f_{n-j}(x) z^j
=\prod_{j=1}^n (z-\mu_j(x)), \quad \{ \mu_j(x)\}_{j=1,\dots,n}
\subset
\bbC, \, x\in\bbC, \lb{2.10}
\end{equation}
equations \eqref{2.2} and \eqref{2.8}, that is,
$f_{n+1,x}(x)=0$,
$x\in\bbC$, imply
\begin{equation}
F_{n,xxx}(z,x) -4 (V(x)-z) F_{n,x}(z,x) -2V_x(x)
 F_n (z,x)=0. \lb{2.11}
\end{equation}
Multiplying \eqref{2.11} by $F_{n}(z,x)$ and
integrating yields
\begin{equation}
(1/2) F_{n,xx}(z,x) F_n(z,x) - (1/4) \ul
F_{n,x}(z,x)^2 - (V(x)-z) F_n(z,x)^2 = R_{2n+1}(z),
\lb{2.12}
\end{equation}
where the integration constant $R_{2n+1}(z)$ is a monic
polynomial in
$z$ of degree $2n+1$, and hence of the form
\begin{equation}
R_{2n+1} (z) = \prod_{m=0}^{2n} (z-E_m),
\quad \{ E_m\}_{m=0,\dots,2n} \subset \bbC.
\lb{2.13}
\end{equation}

Introducing the algebraic eigenspace,
\begin{equation}
\ker (L-z)=\left\{ \psi\colon \bbC\to\bbCinf
\text{  meromorphic}\mid (L -z)\psi=0 \right\}, \quad z\in\bbC
\lb{2.14}
\end{equation}
one verifies
\begin{equation}
P_{2n+1}\big|_{\ker(L-z)} =
\big(F_n (z,x) \f{d}{dx} -(1/2) F_{n,x}(z,x)\big)
\big|_{\ker(L-z)} \lb{2.15}
\end{equation}
and
\begin{align}
\big(P_{2n+1}\big|_{\ker(L-z)}\big)^2
&= \big(-(1/2) F_{n,xx}(z,x)
F_n(z,x) +(1/4) F_{n,x}(z,x)^2 \lb{2.15a} \\
&\quad \,\, + (V(x)-z) F_n(z,x)^2\big)\big|_{\ker(L-z)}
=-R_{2n+1}(z)\big|_{\ker(L-z)}. \no
\end{align}
In agreement with a celebrated result by Burchnall--Chaundy
\cite{BC23},
\cite{BC28}, \cite{BC32} (see also \cite{GG91}, \cite{Pr96},
\cite{Pr98}, \cite{SW85}, \cite{Wi85}), this yields
\begin{equation}
P_{2n+1}^2+R_{2n+1}(L)=0. \lb{2.16}
\end{equation}
Equation \eqref{2.16} naturally leads to the hyperelliptic  curve
$\calK_{n}$ defined by
\begin{equation}
\calK_{n}\colon y^2=R_{2n+1}(z), \quad
R_{2n+1}(z) = \prod_{m=0}^{2n} (z-E_m).
\label{2.17}
\end{equation}
The one-point compactification of $\calK_{n}$ by joining
$P_{\infty}$, the point at infinity, is then denoted by
$\ol {\calK_{n}}$.
A general point $P\in\ol{\calK_{n}}\backslash\{P_{\infty}\}$ will
be  denoted by $P=(z,y)$, where $y^2=R_{2n+1}(z).$

For future purposes we define the involution $*$ on
$\ol{\calK_{n}}$  (i.e., the sheet exchange map) by
\begin{equation}
*\colon\ol{\calK_{n}}\to\ol{\calK_{n}}, \quad
P=(z,y)\mapsto P^{*}=(z,-y), \quad
P_{\infty}^{*}=P_{\infty}. \label{2.19}
\end{equation}

Returning to \eqref{2.10}--\eqref{2.13} and \eqref{2.17},
we introduce the polynomial $H_{n+1}(z,x)$ of degree $n+1$
in $z\in\bbC,$
\begin{equation}
H_{n+1}(z,x)=(1/2){F}_{n,xx}(z,x)-(V(x)-z)
{F}_n(z,x) \lb{2.19a}
\end{equation}
and note that \eqref{2.12} then can be rewritten as
\begin{equation}
{R}_{2n+1}(z)+(1/4)F_{n,x}(z,x)^2=F_n(z,x)
H_{n+1}(z,x), \quad z, x\in\bbC. \lb{2.19b}
\end{equation}
Next we define the following meromorphic function
$\phi(P,x)$ on $\ol{\calK_n}$ by
\begin{align}
\phi (P,x) &= \frac{iy(P) +(1/2) F_{n,x} (z, x)}
{F_n (z,x)} \lb{2.20} \\
&=\frac{H_{n+1}(z,x)}{-iy(P)+(1/2)F_{n,x}(z,x)},
\quad P=(z,y)\in \ol{\calK_n}, \, x\in\bbC, \lb{2.20a}
\end{align}
where $y(P)$ denotes the meromorphic function on
$\ol{\calK_n}$ obtained by solving
$y^2=R_{2n+1}(z)$ with $P=(z,y).$ By \eqref{2.10},
\eqref{2.19b}, and \eqref{2.20a}, the poles of $\phi(P,x)$ on
$\ol {\calK_{n}}$ are given by
\begin{equation}
\hat \mu_j(x)=(\mu_j(x),-(i/2)F_{n,x}(\mu_j(x),x)),
\, 1\leq j\leq n, \text{ and by } P_\infty. \lb{2.20b}
\end{equation}
Given
$\phi(P,x),$ we introduce the meromorphic Baker-Akhiezer function
$\psi(P,x,x_0)$ on $\ol{\calK_n}\backslash\{P_\infty\}$ by
\begin{equation}
\psi(P,x,x_0) =\exp \bigg( \int_{x_0}^x  dx' \,
\phi(P,x')\bigg), \quad  P  = (z,y)\in\ol
{\calK_n}\backslash\{P_\infty\},\, x,x_0 \in\bbC,
\lb{2.21}
\end{equation}
choosing a smooth non-selfintersecting path from $x_{0}$ to $x$
which avoids singularities of $\phi (P,x).$ Straightforward
computations then yield the Riccati-type equation
\begin{equation}
 \phi_x (P,x) + \phi(P,x)^2 = V(x) -z, \lb{2.22}
\end{equation}
for $\phi(P,x)$ and show that $\psi(P,x,x_0)$ satisfies the
Schr\"odinger equation,
\begin{equation}
 (L-z(P)) \psi(P,\dott,x_0) =0. \lb{2.23}
\end{equation}
Moreover, one computes
\begin{align}
\psi (P,x,x_0)\psi (P^*, x,x_0)&= F_n (z,x)
/ F_n (z,x_0),\lb{2.24}\\
W(\psi (P,\dott,x_0), \psi (P^*, \dott, x_0))&=
-i2y(P)/F_n(z,x_0), \lb{2.25}
\end{align}
where $W(f,g)(x)=f(x)g'(x)-f'(x)g(x)$ denotes the Wronskian
of $f$ and $g.$

Given $\psi(P,x,x_0)$ we can define the formal diagonal Green's
function $g(P,x),$ $P\in\ol{\calK_n}\backslash\{P_\infty\},$
of $L$ by
\begin{align}
g(P,x)&=\f{\psi (P,x,x_0) \psi (P^*, x, x_0)}
{W(\psi (P,\dott,x_0), \psi (P^*, \dott, x_0))}
=\frac{iF_n(z,x)}{2y(P)} \lb{2.26} \\
&=\f{i}{2}\f{\prod_{j=1}^n(z-\mu_j(x))}
{\big(\prod_{m=0}^{2n}(z-E_m)\big)^{1/2}}, \quad
 P=(z,y)\in\ol{\calK_n}\backslash\{P_\infty\},
\; x\in\bbC, \lb{2.26a}
\end{align}
and observe that $g(P,x)$ satisfies the universal differential
equations
\begin{equation}
-g_{xxx}(P,x)+4(V(x)-z)g_x(P,x)+2V^{\prime}(x)g(P,x)=0 \lb{2.27}
\end{equation}
and
\begin{equation}
-2g_{xx}(P,x)g(P,x)+g_x(P,x)^2+4(V(x)-z)g(P,x)^2=1 \lb{2.28}
\end{equation}
after inserting \eqref{2.26} into \eqref{2.11} and \eqref{2.12}.
The universal equation \eqref{2.27} was known to Appell in 1880
\cite{Ap80}.

Moreover, comparing terms of order $z^{2n}$ in
\eqref{2.12} yields the trace formula for $V(x),$
\begin{equation}
V(x)=E_0+\sum_{j=1}^n(E_{2j-1}+E_{2j}-2\mu_j(x)). \lb{2.29}
\end{equation}

Next we turn to the time-dependent case. In order to define the
time-dependent KdV hierarchy one introduces
a deformation parameter
$t_r\in\bbC$ into
$V(x),$ that is, one systematically makes the replacements $V(x)\to
V(x,t_r),$
$L\to L(t_r),$ $P_{2r+1}\to P_{2r+1}(t_r),$
$f_j(x)\to f_j(x,t_r),$ $F_n(z,x)\to F_n(z,x,t_r),$
$\mu_j(x)\to\mu_j(x,t_r),$ $\phi(P,x)\to\phi(Px,t_r),$ etc., and
then postulates the Lax commutator representation,
\begin{equation}
0=\f{d}{dt_r}L(t_r) - [P_{2r+1}(t_r),L(t_r)] = V_{t_r}-2
f_{r+1,x}  =\KdV_r(V), \quad r\in\bbN_0. \lb{2.30}
\end{equation}
Varying $r\in\bbN_0$ defines the time-dependent KdV hierarchy,
denoted by $\KdV_r(\cdot).$
Explicitly,  one obtains for the first few KdV equations,
\begin{align}
\KdV_0 (V) &= V_{t_0} - V_x =0, \no\\
\KdV_1 (V) & = V_{t_1} + (1/4) V_{xxx} -
(3/2) VV_x - c_1V_x=0, \lb{2.31} \\
\KdV_2 (V) &= V_{t_2} -(1/16) V_{xxxxx} +(5/8) VV_{xxx}
+(5/4) V_x V_{xx} - (15/8) V^2 V_x \no\\
& \hspace*{4mm}
 - c_2 V_x + c_1 ((1/4) V_{xxx} -(3/2) VV_x)=0,
\text{  etc.,} \no
\end{align}
where, of course, $\KdV_1 (V)$ is {\it the} KdV equation.

In terms of $F_n(z,x,t_r),$ \eqref{2.30} can be rewritten as
\begin{equation}
V_{t_r}(x,t_r)=-(1/2)F_{n,xxx}(z,x,t_r)+2(V(x)-z)
F_{n,x}(z,x,t_r)+V_x(x)F_n(z,x,t_r). \lb{2.32}
\end{equation}

The time-dependent Baker-Akhiezer function (i.e., the analog of
\eqref{2.21}) is defined by
\begin{align}
&\psi(P,x,x_0,t_r,t_{0,r})=\exp\bigg(\int_{x_0}^x dx^\prime \,
\phi(P,x^\prime,t_r)+\int_{t_{0,r}}^{t_r}ds\,
({\wti F}_r (z,x_0,s)\phi(P,x_0,s) \no \\
&\hspace*{1cm} -(1/2){\wti F}_{r,x}  (z,x_0,s)) \bigg),
\quad P=(z,y)\in\ol {\calK_n}\backslash\{P_\infty\},\,
x,x_0,t_r,t_{0,r} \in\bbC \lb{2.32a}
\end{align}
with $\phi(P,x,t_r)$ given as in \eqref{2.20},
\eqref{2.20a} (replacing $F_n(z,x)$
 by $F_n(z,x,t_r)$ and $H_{n+1}(z,x)$ by
$H_{n+1}(z,x,t_r),$ respectively) and one verifies that
$\psi(\cdot,x,x_0,t_r,t_{0,r})$ is meromorphic on
$\calK_n,$ $(L(t_r)-z(P))\psi(P,\cdot,x_0,t_r,t_{0,r})
=0,$ and (cf.~\cite[Ch.~1]{GH00},
\cite{GRT96}),
\begin{align}
&\phi_{t_r}(P,x,t_r)=\partial_x({\wti F}_r (z,x,t_r)
\phi(P,x,t_r)-(1/2){\wti F}_{r,x} (z,x,t_r)), \lb{2.32b} \\
&\psi_{t_r}(P,x,x_0,t_r,t_{0,r})=(\wti P_{2r+1}(t_r)\psi)
(P,x,x_0,t_r,t_{0,r}) \lb{2.32c} \\
&={\wti F}_r (z,x,t_r)\psi_x(P,x,x_0,t_r,t_{0,r})
-(1/2){\wti F}_{r,x}
(z,x,t_r)\psi(P,x,x_0,t_r,t_{0,r}), \no \\
&F_{n,t_r} (z,x,t_r)={\wti F}_r (z,x,t_r)F_{n,x}
(z,x,t_r)  -{\wti F}_{r,x} (z,x,t_r)F_n (z,x,t_r).
\lb{2.32d}
\end{align}

Finally, we briefly consider the algebro-geometric initial value
problem. Fix $\Omega\subset\bbC^2,$ $\Omega$ open,
$t_{0,r}\in\bbC,$ then on the level of Lax differential
expressions this initial value problem reads
\begin{align}
[P_{2n+1}(t_{0,r}),L(t_{0,r})]&=0, \quad (x,t_{0,r})\in\Omega,
\lb{2.33} \\
\f{d}{dt_{r}}L(t_{r})-[{\wti P}_{2r+1}(t_r),L(t_r)]&=0,
\quad (x,t_{r})\in\Omega, \lb{2.34}
\end{align}
where the integration constants $(c_1,\dots,c_n)$ in $P_{2n+1}$
and those in $\wti P_{2r+1},$ denoted by $(\ti c_1,\dots,\ti
c_r),$ are independent of each other. Similarly, we denote by
${\ti f}_j,$ $1\leq j\leq r,$ ${\ti f}_{r+1,x}$
${\wti F}_r,$
$\wti \KdV_r(\cdot),$ the corresponding quantities obtained from
$f_j,$ $1\leq j\leq r,$ $f_{r+1,x}$ $F_r,$
$\KdV_r(\cdot),$ replacing the set of integration constants
$(c_1,\dots,c_n)$ by $(\ti c_1,\dots,\ti c_r).$ Since KdV
flows are
expected to be isospectral we next replace \eqref{2.33},
\eqref{2.34} by
\begin{align}
[P_{2n+1}(t_{r}),L(t_{r})]&=0, \quad (x,t_{r})\in\Omega,
\lb{2.35} \\
\f{d}{dt_{r}}L(t_{r})-[{\wti P}_{2r+1}(t_r),L(t_r)]&=0,
\quad (x,t_{r})\in\Omega. \lb{2.36}
\end{align}
In terms of the polynomial formalism, \eqref{2.35}
and \eqref{2.36} become ($(x,t_r)\in\Omega$)
\begin{align}
&(1/2)F_{n,xxx}(0,x,t_r) -2V(x)F_{n,x}(0,x,t_r)
-V_x(x,t_r)F_n(0,x,t_r)=0, \lb{2.37} \\
&V_{t_r}(x,t_r)=-\tfrac{1}{2}{\wti
F}_{r,xxx}(0,x,t_r)  +2V(x){\wti F}_{r,x}(0,x,t_r)
 +V_x(x,t_r){\wti F}_r (0,x,t_r). \lb{2.38}
\end{align}
\eqref{2.37} and \eqref{2.38} are of course equivalent to
\begin{align}
\sKdV_n (V)&=-2f_{n+1,x}=0, \quad (x,t_r)\in\Omega,
\lb{2.39} \\
\wti \KdV_r (V)&=V_{t_r}-2{\ti f}_{r+1,x}=0,
\quad (x,t_r)\in\Omega. \lb{2.40}
\end{align}
Hence $V(x,t_r),$ for each fixed $t_r$ (with $(x,t_r)\in\Omega$)
satisfies the stationary $n$th KdV equation, while $V(x,t_r),$
$(x,t_r)\in\Omega,$ satisfies the $r$th time-dependent KdV
equation.

%%%%%%%%%%%%%%%%%%%%%%%%%%%%%%%%%%%%%%%%%%%%%%%%%%%%%%%%%%%%%%
\begin{remark} \lb{r2.1} We emphasize that in the general case
where
$V(x,t_r)$ is complex-valued, \eqref{2.39}, \eqref{2.40} does not
necessarily have solutions for any distribution of
$\{E_m\}_{m=0,\dots,2n}\subset\bbC$ as demonstrated by Birnir
\cite{Bi86b}. Unique solvability of the isospectral deformation
problem
\eqref{2.39}, \eqref{2.40} (or equivalently, \eqref{2.37},
\eqref{2.38}) can be reduced to unique solvability of the
standard  first-order
system of Dubrovin equations (see, e.g., \cite{DT76},
\cite[Sect.~12.3]{Di91}, \cite{Du75}, \cite{GD79}, \cite{GH98},
\cite{GRT96}, \cite[Chs.~10, 12]{Le87}, \cite[Ch.~4]{Ma86}).
As long as the curve $\calK_n$ is nonsingular (i.e.,
$E_m\neq E_{m^\prime}$ for $m\neq m^\prime$) and none of the
$\mu_j(x,t_r)$ collide (i.e.,
$\mu_j(x,t_r)\neq\mu_{j^\prime}(x,t_r)$ for $(x,t_r)\in\Omega,$
$\Omega\in\bbC^2$ open), the Dubrovin system is uniquely solvable
on $\calK_n$ and then $V(x,t_r),$
$(x,t_r)\in\Omega,$ is recovered from the time-dependent trace
formula
\begin{equation}
V(x,t_r)=E_0+\sum_{j=1}^n(E_{2j-1}+E_{2j}-2\mu_j(x,t_r)).
\lb{2.48}
\end{equation}
The corresponding $F_n(z,x,t_r),$
${\wti F}_r(z,x,t_r),$ $(x,t_r)\in\Omega,$ then satisfy the
isospectral deformation  equations \eqref{2.37}, \eqref{2.38}.
The case of collisions between the $\mu_j(x,t_r)$ is more subtle
but can also be treated as shown by Birnir
\cite{Bi86a}, \cite{Bi86b}.
In particular, in the case of nonsingular curves
$\calK_n$ with real branch points
$\{E_m\}_{m=0,\dots,2n}\subset\bbR,$ $E_m\neq E_{m^\prime}$ for
$m\neq m^\prime$ no collisions between any of the
$\mu_j(x,t_r)$ for $(x,t_r)\in\bbR^2$ can ever occur due to the
strict separation of all spectral gaps by spectral bands. In this
case, \eqref{2.39}, \eqref{2.40} yield a smooth solution
$V(x,t_r),$ $(x,t_r)\in\bbR^2,$ expressable in terms of the
Riemann theta function
$\theta(\ul z)$ associated with $\ol{\calK_n}$ by the
Its-Matveev  formula \cite{IM75}
\begin{equation}
V(x,t_r)=\Lambda_0-2\partial^2_x\ln(\theta (\ul A+\ul Bx
+\ul C_rt_r)) \lb{2.41}
\end{equation}
for appropriate $\Lambda_0\in\bbR,$ $\ul A,\ul B,
\ul C_r\in\bbR^n,$ depending on the homology basis chosen on
$\ol {\calK_n}.$
\end{remark}
%%%%%%%%%%%%%%%%%%%%%%%%%%%%%%%%%%%%%%%%%%%%%%%%%%%%%%%%%%%%%%

%%%%%%%%%%%%%%%%%%%%%%%%%%%%%%%%%%%%%%%%%%%%%
\section{The infinite genus limit} \lb{s3}
%%%%%%%%%%%%%%%%%%%%%%%%%%%%%%%%%%%%%%%%%%%%%

In this section we will extend the approach of Section~\ref{s2}
to a special class of infinite genus curves whose branch points
accumulate at $+\infty.$ In particular, we will explore the
extent to which the Burchnall-Chaundy results generalize to the
transcendental hyperelliptic curves at hand.

We start with the stationary (i.e., time-independent) case. Our
point of departure is a rewriting of the diagonal Green's function
$g(P,x)$ in the finite genus case
\eqref{2.26a} in the form
\begin{equation}
g(P,x)=\f{1}{2}\f{\prod_{j=1}^n
((\mu_j(x)-z)/(E_{2j-1}E_{2j})^{1/2})}
{\big((E_0-z)\prod_{m=1}^{2n}(1-(z/E_m))\big)^{1/2}}=
\f{\uF_n(z,x)}{2y(P)}, \lb{3.1}
\end{equation}
(for $P=(z,y)\in\uK_n$) where
\begin{equation}
\uF_n(z,x)=\prod_{j=1}^n ((\mu_j(x)-z)/(E_{2j-1}E_{2j})^{1/2})
\lb{3.2}
\end{equation}
and $\uK_n$ denotes the hyperelliptic curve
\begin{equation}
\uK_n: y^2=\uR_{2n+1}(z), \quad \uR_{2n+1}(z)=(E_0-z)
\prod_{m=1}^{2n} (1-(z/E_m)). \lb{3.3}
\end{equation}

For simplicity we assume that $E_m\neq 0,$ $m=1,\dots,2n$ and
that $\uR_{2n+1}(z)$ has a simple zero at $z=E_0.$ According to  our
convention in Section~\ref{s2}, the products in
\eqref{3.1}  repeat each factor associated with $\mu_j(x)$ and
$E_m$ according to their multiplicities. Introducing the following
Hypothesis~\ref{h3.1} will then enable us to treat the case
$n=\infty$ in \eqref{3.1}--\eqref{3.3}.
%%%%%%%%%%%%%%%%%%%%%%%%%%%%%%%%%%%%%%%%%%%%%%%%%%%%%%%%%%%%%%%%%%
\begin{hypothesis} \lb{h3.1}
Let $\calI\subseteq\bbR$ be open and $x\in\calI.$ \\
\noindent {\rm (i)} Assume $\{E_m\}_{m\in\bbN_0}\subset\bbC$
such that
$\sum_{m\in\bbN} |E_m|^{-1}<\infty,$
$\Re(E_m)\underset{m\to\infty}{\longrightarrow} +\infty,$
$\Im(E_m)\underset{m\to\infty}{\longrightarrow} 0,$ ordering the
$E_m$ according to their absolute values counting multiplicity. \\
\noindent {\rm (ii)} Suppose
$\{\mu_j(x)\}_{j\in\bbN}\subset\bbC$ such that
$\{1-\mu_j(x)(E_{2j-1}E_{2j})^{-1/2}\}_{j\in\bbN}
\in \ell^1(\bbN)$ for all $x\in\calI,$ where we agree to take the
principal  branch of $(E_{2j-1}E_{2j})^{1/2}$ (such that
$\arg((E_{2j-1}E_{2j})^{1/2})
\underset{j\to\infty}{\longrightarrow} 0$). \\
\noindent {\rm (iii)} For all $x\in\calI,$ assume
$\{E_{2j-1}+E_{2j}-2\mu_j(x)\}_{j\in\bbN}\in \ell^1(\bbN)$
and $\{(E_{2j-1}E_{2j}-\mu_j(x)^2)/E_{2j}\}_{j\in\bbN}\in
\ell^1(\bbN).$
\end{hypothesis}
%%%%%%%%%%%%%%%%%%%%%%%%%%%%%%%%%%%%%%%%%%%%%%%%%%%%%%%%%%%%%%%%%%

Hypothesis~\ref{h3.1}\,(i) and the Cauchy-Schwarz
inequality then yield
$\sum_{j\in\bbN} |E_{2j-1}\\E_{2j}|^{-1/2} \leq
(\sum_{j\in\bbN}|E_{2j-1}|^{-1})^{1/2}
(\sum_{j\in\bbN} |E_{2j}|^{-1})^{1/2}<\infty.$
Thus, assuming Hypotheses~\ref{h3.1}\,{\rm (i), (ii)} for the
remainder of  this paper, one can define the transcendental
(infinite genus) hyperelliptic curve
\begin{equation}
\calK_\infty: y^2=R_\infty(z), \quad R_\infty(z)= (E_0-z)
\prod_{m\in\bbN} (1-(z/E_m)). \lb{3.9}
\end{equation}
In analogy to Section~\ref{s2}, points
$P\in\calK_{\infty}$ are denoted by
$P=(z,y)$, where $y^2=R_{\infty}(z)$ and the involution $*$ on
$\calK_{\infty}$ is defined as in \eqref{2.19} by
\begin{equation}
*\colon\calK_{\infty}\to\calK_{\infty}, \quad
P=(z,y)\mapsto P^{*}=(z,-y). \label{3.9a}
\end{equation}
In addition we define the diagonal Green's function,
\begin{equation}
g(P,x)=\f{1}{2}\f{\prod_{j\in\bbN}
((\mu_j(x)-z)/(E_{2j-1}E_{2j})^{1/2})}{\big((E_0-z)\prod_{m\in\bbN}
(1-(z/E_m))\big)^{1/2}}=\f{F_\infty (z,x)}{2y(P)} \lb{3.7}
\end{equation}
(for $P=(z,y)\in\calK_\infty$), with absolutely convergent infinite
products in \eqref{3.9} and \eqref{3.7}. Here
\begin{equation}
F_\infty (z,x)=\prod_{j\in\bbN}
((\mu_j(x)-z)/(E_{2j-1}E_{2j})^{1/2}), \lb{3.8}
\end{equation}
and as in \eqref{3.2} and \eqref{3.3} we asume that in the infinite
products
\eqref{3.9} and \eqref{3.8} each factor associated with $\mu_j(x)$
and
$E_m$ is repeated according to its multiplicity. For simplicity
we suppose again that $R_\infty(z)$ has a simple zero at $E_0.$
This can always be achieved by a simple translation $V(x)\to
V(x)+z_0$ for some $z_0\in\bbC.$

Taking \eqref{3.7}--\eqref{3.9} as the model for diagonal Green's
functions of differential expressions $L=-\f{d^2}{dx^2}+V(x)$
associated with infinite genus curves of the type \eqref{3.9}, we
now postulate the following hypothesis.

%%%%%%%%%%%%%%%%%%%%%%%%%%%%%%%%%%%%%%%%%%%%%%%%%%%%%%%%%%%%%%%%%%
\begin{hypothesis} \lb{h3.2}
Assume Hypothesis~\ref{h3.1}\,(i), (ii) and suppose $x\in\calI.$
Define
$\calK_\infty,$ $g(P,x),$ $P\in\calK_\infty,$ $F_\infty (z,x),$
and $R_\infty (z)$ as in \eqref{3.7}--\eqref{3.9}. Denote by
$C_\varepsilon=\{z\in\bbC\,|\, \Re(z)>0,\,|\arg(z)|<\varepsilon\}$
the cone along the positive real axis with apex at the
origin and opening angle $\varepsilon,$ $0<\varepsilon<\pi/2.$ \\
\noindent {\rm (i)} Suppose that $V\in C^1(\calI)$ and
$g(P,\cdot)\in C^3(\calI)$ for all $P\in\calK_\infty.$ \\
\noindent {\rm (ii)} Assume that $g(P,x)$ satisfies
\begin{equation}
-2g_{xx}(P,x)g(P,x)+g_x(P,x)^2 + 4(V(x)-z)g(P,x)^2 =1. \lb{3.10}
\end{equation}
\noindent {\rm (iii)} Suppose that for each $x\in\calI,$
\begin{equation}
g(P,x)=(i/2)z(P)^{-1/2} +(i/4)z(P)^{-3/2}V(x) +o(z(P)^{-3/2})
\lb{3.10b}
\end{equation}
as $z\to \infty,$ $z\in\bbC\backslash C_\varepsilon.$ \\
\noindent {\rm (iv)} Assume that as $z\to \infty,$
$z\in\bbC\backslash C_\varepsilon,$
\begin{align}
g(P,x)&=(i/2)z(P)^{-1/2} +o(z(P)^{-1/2}), \lb{3.10c} \\
g_x (P,x)&=o(z(P)^{-1/2}), \lb{3.10d}
\end{align}
uniformly with respect to $x$ as long as $x$ varies in compact
subintervals of $\calI.$
\end{hypothesis}
%%%%%%%%%%%%%%%%%%%%%%%%%%%%%%%%%%%%%%%%%%%%%%%%%%%%%%%%%%%%%%%%%%
\noindent For the time being we prefer to develop the theory for
$x\in\calI$ rather than $x\in\bbR$ (or $x\in\bbC$) in order to
include a possible meromorphic behavior of $V(x).$ Moreover,
we are not necessarily restricting ourselves to smooth
functions $V$ (such  as $V\in C^\infty (\calI),$ etc.).

Differentiating \eqref{3.10} with respect to $x$ yields the linear
third-order differential equation
\begin{equation}
-g_{xxx}(P,x)+4(V(x)-z)g_x(P,x)+2V^\prime (x)g(P,x)=0, \quad
P\in\calK_\infty, \, x\in\calI. \lb{3.10a}
\end{equation}

Given Hypothesis~\ref{h3.2}, we introduce the entire function
$H_\infty(z,x)$ with respect to $z$, by
\begin{equation}
H_\infty(z,x)=(1/2)F_{\infty,xx}(z,x)+(z-V(x))F_\infty(z,x),
\quad z\in\bbC, \, x\in\calI. \label{3.11}
\end{equation}
Inserting $g(P,x)=F_\infty (z,x)/(2y(P))$ (cf. \eqref{3.7} and
\eqref{3.9}) into \eqref{3.10} and \eqref{3.10a} then yields
($z\in\bbC, \, x\in\calI$)
\begin{align}
&-\tfrac{1}{2}F_{\infty,xx}(z,x)F_\infty (z,x)+(V(x)-z)F_\infty
(z,x)^2  + \tfrac{1}{4}F_{\infty,x}(z,x)^2=R_\infty (z),  \lb{3.12}
\end{align}
and
\begin{equation}
-F_{\infty,xxx}(z,x)+4(V(x)-z)F_{\infty,x}(z,x)
+2V^{\prime}(x)F_\infty
(z,x)=0. \lb{3.12a}
\end{equation}
Combining \eqref{3.11} and \eqref{3.12} results in
\begin{equation}
R_\infty(z)+(1/4) F_{\infty,x}(z,x)^2 =F_\infty(z,x)H_\infty(z,x),
\quad z\in\bbC, \, x\in\calI.	\label{3.13}
\end{equation}
Thus, we may define the following fundamental meromorphic function
$\phi(P,x)$ on $\calK_\infty$,
\begin{align}
\phi (P,x) &= \frac{-y(P) +(1/2) F_{\infty,x} (z, x)}
{F_\infty (z,x)} \lb{3.14} \\
 & = \frac{H_\infty (z,x)}{y(P) +(1/2) F_{\infty,x}(z, x)},
\quad   P  = (z,y)\in\calK_\infty, \; x\in\calI, \lb{3.15}
\end{align}
where $y(P)$ denotes the analytic function on $\calK_\infty$
obtained by solving $y^2=R_\infty(z)$ with $P=(z,y)$
(cf.~\eqref{3.9}). By \eqref{3.8}, \eqref{3.13}, and \eqref{3.15}
we can identify the pole positions
$\{\hat \mu_j(x)\}_{j\in\bbN}$ of $\phi(P,x)$ on $\calK_\infty$
as
\begin{equation}
\hat \mu_j(x)=(\mu_j(x),-(1/2)F_{\infty,x}(\mu_j(x),x)), \quad
j\in\bbN. \lb{3.15a}
\end{equation}
In addition, we introduce the analog of the
stationary Baker-Akhiezer function
$\psi(P, x, x_0)$ on $\calK_\infty$ by
\begin{equation}
\psi(P,x,x_0) =\exp \bigg( \int_{x_0}^x  dx' \,
\phi(P,x')\bigg), \quad  P  = (z,y)\in\calK_\infty,\,
x,x_0 \in\calI, \lb{3.16}
\end{equation}
choosing a smooth non-selfintersecting path from $x_{0}$ to $x$
which avoids singularities of $\phi (P,x).$

%%%%%%%lemma%%%%%%%%%%%%%%%%%%%%%%%%%%%%%%%%%%%%%%%%%%%%%%%%%%%%%
\begin{lemma} \lb{l3.3} Assume Hypothesis~\ref{h3.2}\,(i),(ii) and
 let $P= (z,y) \in \calK_\infty,$ $x,x_0\in\calI.$ Then
$\phi(P,x)$ satisfies
\begin{align}
 \phi_x (P,x) + \phi(P,x)^2 &= V(x) -z, \lb{3.17} \\
 \phi(P,x) \phi (P^*, x)&= H_\infty (z,x)
/ F_\infty (z,x),\lb{3.18}\\
\phi (P,x) + \phi (P^*, x)&= F_{\infty,x} (z,x)
/ F_\infty (z,x),\lb{3.19}\\
 \phi(P,x)-\phi (P^*, x)&=-2y(P)/F_\infty (z,x).
\lb{3.20}
\end{align}
$\psi (\cdot,x,x_0)$ is meromorphic on $\calK_\infty$ and
satisfies
\begin{align}
 (L-z(P)) \psi(P,\dott,x_0) &=0, \quad
L=-\f{d^2}{dx^2}+V(x), \lb{3.22} \\
\psi (P,x,x_0) \psi (P^*, x, x_0)
&= F_\infty (z,x) / F_\infty(z,x_0),
\lb{3.23}\\
\psi_x (P,x,x_0) \psi_x (P^*, x, x_0)
&= H_\infty (z,x)/ F_\infty (z,x_0),
\lb{3.24}\\
\psi(P,x,x_0)\psi_x(P^*,x,x_0)
+&\psi(P^*,x,x_0)\psi_x(P,x,x_0)
=F_{\infty,x}(z,x)/F_\infty(z,x_0), \lb{3.25} \\
W(\psi (P,\dott,x_0), \psi (P^*, \dott, x_0))&=
2y(P)/F_\infty (z,x_0).
\lb{3.26}
\end{align}
Moreover, one obtains
\begin{equation}
g(P,x)=\frac{F_\infty(z,x)}{2y(P)}
=\f{\psi (P,x,x_0) \psi (P^*, x, x_0)}
{W(\psi (P,\dott,x_0), \psi(P^*, \dott, x_0))}. \lb{3.28}
\end{equation}
\end{lemma}
%%%%%%% end of lemma %%%%%%%%%%%%%%%%%%%%%%%%%%%%%%%%%%%%%%%%%
\begin{proof}
\eqref{3.12} and \eqref{3.14} immediately yield \eqref{3.17}
and \eqref{3.18}.
\eqref{3.19} and \eqref{3.20} are clear from \eqref{3.14}.
\eqref{3.16} and \eqref{3.17} yield \eqref{3.22}, and
\eqref{3.16} and \eqref{3.19} prove \eqref{3.23}.
$\phi=\psi_x/\psi$ and \eqref{3.18}, \eqref{3.23} then yield
\eqref{3.24}--\eqref{3.26}. \eqref{3.28} follows from
\eqref{3.23} and \eqref{3.26}. Finally, by \eqref{3.16},
$\psi(\cdot,x,x_0)$  is clearly meromorphic on
$\calK_\infty$ away from the poles
$\hat \mu_j(x^{\prime})$ of $\phi(\cdot,x^{\prime}).$ By
\eqref{3.14} and \eqref{3.15a} one concludes
\begin{equation}
\phi(P,x^{\prime})\underset{P\to\hat\mu_j(x^{\prime})}{=}
\f{d}{dx^{\prime}}\ln(F_\infty(z,x^{\prime})) + O(1) \text{ as }
z\to \mu_j(x^{\prime}), \quad j\in\bbN, \lb{3.28a}
\end{equation}
and hence $\psi(\cdot,x,x_0)$  is meromorphic on
$\calK_\infty.$
\end{proof}

Next, we derive a trace formula representation of $V(x)$ in
terms of $\{\mu_j(x)\}_{j\in\bbN}.$

%%%%%%%lemma%%%%%%%%%%%%%%%%%%%%%%%%%%%%%%%%%%%%%%%%%%%%%%%%%%%%%
\begin{lemma} \lb{l3.4} Assume Hypotheses~\ref{h3.1} and
\ref{h3.2}\,(i)--(iii). Then
\begin{equation}
V(x)=E_0+\sum_{j\in\bbN}(E_{2j-1}+E_{2j}-2\mu_j(x)), \quad
x\in\calI. \lb{3.29}
\end{equation}
\end{lemma}
%%%%%%%%%%%%%%%%%%%%%%%%%%%%%%%%%%%%%%%%%%%%%%%%%%%%%%%%%%%%%%%%%
\begin{proof}
By Hypothesis~\ref{h3.1}\,{\rm (i)--(iii)} one infers for
$x\in\calI,$
\begin{align}
&\ln\Big(\Big(\prod_{j\in\bbN}(\mu_j(x)-z)^2(E_{2j-1}E_{2j})^{-1}
\Big/\prod_{m\in\bbN}(1-(z/E_m))\Big)^{1/2}\Big) \no \\
&=(1/2)\sum_{j\in\bbN}\ln\bigg(1+\f{E_{2j-1}+E_{2j}-2\mu_j(x)}
{z(1-(E_{2j-1}/z))(1-(E_{2j}/z))} \no \\
&\hspace*{2.72cm} +\f{\mu_j(x)^2-E_{2j-1}E_{2j}}
{z^2(1-(E_{2j-1}/z))(1-(E_{2j}/z))}\bigg) \no \\
&=(1/2)z^{-1}\sum_{j\in\bbN}(E_{2j-1}+E_{2j}-2\mu_j(x))+o(z^{-1})
\text{ as } z\to \infty, \, z\in\bbC\backslash
C_\varepsilon. \lb{3.35a}
\end{align}
In connection with the $o(z^{-1})$ term in \eqref{3.35a} we used
$[\mu_j(x)^2-E_{2j-1}E_{2j}]/[z^2(1-(E_{2j-1}/z))(1-(E_{2j}/z))]=
z^{-1}[\mu_j(x)^2-E_{2j-1}E_{2j}]/[(1-(E_{2j-1}/z))
(z-E_{2j})],$ Hypothesis~\ref{h3.2}\,(iii), and the dominated
convergence theorem (for discrete measures). Together with
\begin{align}
&g(P,x)=\Big(\f{1}{4(E_0-z)}\prod_{j\in\bbN}
(\mu_j(x)-z)^2(E_{2j-1}E_{2j})^{-1}
\Big/\prod_{m\in\bbN}(1-(z/E_m))\Big)^{1/2}, \no \\
&\hspace*{9.3cm} P=(z,y), \lb{3.35b}
\end{align}
\eqref{3.35a} yields
\begin{align}
g(P,x)&=(i/2)z(P)^{-1/2} +(i/4)z(P)^{-3/2}\bigg(E_0+
\sum_{j\in\bbN}(E_{2j-1}+E_{2j}-2\mu_j(x)\bigg) \no \\
&\quad \, +o(z(P)^{-3/2}) \text{ as } z\to \infty, \,
z\in\bbC\backslash  C_\varepsilon \lb{3.35c}
\end{align}
and hence \eqref{3.29} comparing \eqref{3.35c} and \eqref{3.10b}.
\end{proof}
%%%%%%%%%%%%%%%%%%%%%%%%%%%%%%%%%%%%%%%%%%%%%%%%%%%%%%%%%%%%%%%%%

The trace formula \eqref{3.29} for $V(x),$ although
well-known in the real-valued (infinite genus) periodic case
and in the algebro-geometric
complex-valued case, appears to be new in the present context
of general complex-valued potentials associated with infinite
genus curves.

%%%%%%%lemma%%%%%%%%%%%%%%%%%%%%%%%%%%%%%%%%%%%%%%%%%%%%%%%%%%%%%
\begin{lemma} \lb{l3.4a} Assume Hypotheses~\ref{h3.1} and
\ref{h3.2}. In addition, let $x, x_0\in\calI,$
$P=(z,y)\in\calK_\infty,$ $z\in\bbC\backslash
C_\varepsilon.$ Then, as $z\to \infty,$ $z\in\bbC\backslash
C_\varepsilon,$
\begin{equation}
\phi(P,x)=iz(P)^{1/2}+o(z(P)^{1/2}), \lb{3.21}
\end{equation}
uniformly with respect to $x$ as long as $x$ varies in compact
subintervals of $\calI.$
\end{lemma}
%%%%%%%%%%%%%%%%%%%%%%%%%%%%%%%%%%%%%%%%%%%%%%%%%%%%%%%%%%%%%%%%%
\begin{proof}
It suffices to combine \eqref{3.7}, \eqref{3.10c}, \eqref{3.10d},
and \eqref{3.14}.
\end{proof}
%%%%%%%%%%%%%%%%%%%%%%%%%%%%%%%%%%%%%%%%%%%%%%%%%%%%%%%%%%%%%%%%%

Next we state the following result.

%%%%%%%lemma%%%%%%%%%%%%%%%%%%%%%%%%%%%%%%%%%%%%%%%%%%%%%%%%%%%%%
\begin{lemma} \lb{l3.5} Assume Hypothesis~\ref{h3.2}\,(i), (ii)
and let $P=(z,y)\in\calK_\infty.$ Then
\begin{align}
&\big[g(P,x)\f{d}{dx}-(1/2)g_x(P,x),L\big]\big|_{\ker(L-z)}=0,
\lb{3.30}\\
&\big((g(P,x)\f{d}{dx}-(1/2)g_x(P,x))\big|_{\ker(L-z)}\big)^2
=(1/4)\big|_{\ker(L-z)}. \lb{3.31}
\end{align}
\end{lemma}
%%%%%%%%%%%%%%%%%%%%%%%%%%%%%%%%%%%%%%%%%%%%%%%%%%%%%%%%%%%%%%%%%
\begin{proof}
\eqref{3.30} is a direct consequence of \eqref{3.10a},
and \eqref{3.10} immediately implies \eqref{3.31}.
\end{proof}
%%%%%%%%%%%%%%%%%%%%%%%%%%%%%%%%%%%%%%%%%%%%%%%%%%%%%%%%%%%%%%%%%

Equations \eqref{3.30} and \eqref{3.31} are of course the analogs
of
\begin{align}
&\big[\uF_n(z,x)\f{d}{dx}-(1/2)\uF_{n,x}(z,x),L\big]\big|_{\ker(L-z)}=0,
\lb{3.32}\\
&\big((\uF_n(z,x)\f{d}{dx}-(1/2)\uF_{n,x}(z,x))\big|_{\ker(L-z)}\big)^2
=\uR_{2n+1}(z)\big|_{\ker(L-z)} \lb{3.33}
\end{align}
(cf.~\eqref{2.8}, \eqref{2.15}, \eqref{2.15a}), which result
from restriction of
\begin{align}
&[\uP_{2n+1},L]=0, \lb{3.34} \\
&\uP_{2n+1}^2=-\uR_{2n+1} (L) \lb{3.35}
\end{align}
to $\ker(L-z),$ with
\begin{align}
&\uP_{2n+1}=\uF_n(L,x)\f{d}{dx}-(1/2)\uF_{n,x}(L,x), \lb{3.36} \\
&\uP_{2n+1}\big|_{\ker(L-z)}=\big(\uF_n(z,x)\f{d}{dx}
-(1/2)\uF_{n,x}(z,x)\big)\big|_{\ker(L-z)} \lb{3.37}
\end{align}
(cf.~\eqref{2.8}, \eqref{2.16}). Thus, denoting
$\calJ=\{1,\dots,n\},$ in the case of finite genus $n$ and
$\calJ=\bbN$ in the infinite genus  case, and by $\#(J)$ the
cardinality of $\calJ,$ the universal differential  equation
\eqref{3.10} for $g(P,x)$ together with the  (reflectionless) ansatz
\begin{equation}
g(P,x)=\f{1}{2}\f{\prod_{j=1}^{\#(\calJ)}
((\mu_j(x)-z)/(E_{2j-1}E_{2j})^{1/2})}
{\big((E_0-z)\prod_{m=1}^{2\#(\calJ)}(1-(z/E_m))\big)^{1/2}},
\quad P=(z,y)\in\calK_{\#(\calJ)}, \lb{3.38}
\end{equation}
then leads to algebraically integrable systems associated
with hyperelliptic curves $\calK_{\#(J)}:
y^2=(E_0-z)\prod_{m=1}^{2\#(\calJ)} (1-(z/E_m))$ of (arithmetic)
genus $\#(J)$ and thus unifies the finite and infinite genus
cases at hand. In particular, introducing the infinite-order
differential expression
\begin{align}
&P_\infty=F_\infty(L,x)(d/dx)-(1/2)F_{\infty,x}(L,x), \lb{3.40} \\
&P_\infty\big|_{\ker(L-z)}=\big(F_\infty(z,x)(d/dx)
-(1/2)F_{\infty,x}(z,x)\big)\big|_{\ker(L-z)}, \lb{3.41}
\end{align}
Lemma~\ref{l3.5} implies
\begin{align}
&\big[P_{\infty},L\big]\big|_{\ker(L-z)} \lb{3.42} \\
&=\big(-(1/2)F_{\infty,xxx}(z,x)+2(V(x)-z)F_{\infty,x}(z,x)
+V^\prime (x)F_\infty(z,x)\big)\big|_{\ker(L-z)}=0, \no \\
&\big(P_{\infty}\big|_{\ker(L-z)}\big)^2
=R_{\infty}(L)\big|_{\ker(L-z)}, \lb{3.43}
\end{align}
with $F_\infty (z,x)$ and $R_\infty (z)$ defined in \eqref{3.8}
and \eqref{3.9}. In analogy to \eqref{2.8}, \eqref{2.10},
and \eqref{2.11} we thus define the  stationary $\sKdV_\infty$
equation by
\begin{equation}
\sKdV_\infty(V)=(1/2)F_{\infty,xxx}(0,x)
-2V(x)F_{\infty,x}(0,x)-V^\prime (x)F_\infty(0,x)=0. \lb{3.44}
\end{equation}

%%%%%%%%%%%%%%%%%%%%%%%%%%%%%%%%%%%%%%%%%%%%%%%%%%%%%%%%%%%%%%
\begin{remark} \lb{r3.6}
We emphasize that the formalism in this section applies to general
periodic (complex-valued) potentials $V(x)$ satisfying $V\in
C^1(\bbR).$ Let $\omega >0$ denote the period of
$V.$ By  standard Floquet theory one introduces a fundamental
system of solutions $s(z,x,x_0),$ $c(z,x,x_0)$ of
$-\psi^{''}(z,x) +(V(x)-z)\psi(z,x)=0$ defined by
\begin{align}
s(z,x_0,x_0)&=0, \quad s_x(z,x_0,x_0)=1, \lb{3.44a} \\
c(z,x_0,x_0)&=1, \quad c_x(z,x_0,x_0)=0, \quad z\in\bbC, \,
x, x_0 \in \bbR. \lb{3.44b}
\end{align}
The Floquet discriminant $\Delta(z)$ is then given by
\begin{equation}
\Delta(z)=(c(z,x_0+\omega,x_0)+s_x(z,x_0+\omega,x_0))/2,
\quad z\in\bbC \lb{3.44c}
\end{equation}
and one verifies (cf.~\cite[Sect.~3.4]{Ma86},
\cite{ST96})
\begin{align}
&\Delta(z)^2-1=\omega^2(E_0-z)\prod_{m\in\bbN}\big((E_{2m-1}-z)
(E_{2m}-z)(\omega/\pi)^4m^{-4}\big), \lb{3.44d} \\
&E_{\substack{2m-1\\ 2m}}=(m\pi\omega^{-1}+c_1 m^{-1}
+c_2 m^{-2}\mp \delta_m m^{-2}+\varepsilon_m^\mp m^{-3})^2,
\lb{3.44e} \\
&s(z,x+\omega,x)=\omega\prod_{j\in\bbN}\big((\mu_j(x)-z)
(\omega/\pi)^2j^{-2}\big), \lb{3.44f} \\
&\mu_j(x)=(m\pi\omega^{-1}+c_1 j^{-1}
+c_2 j^{-2}+s_j(x) j^{-2})^2, \lb{3.44g}
\end{align}
where $\{\delta_m\}_{m\in\bbN},$
$\{\varepsilon^\mp_m\}_{m\in\bbN},$
$\{s_m (x)\}_{m\in\bbN} \in \ell^2(\bbN).$ This then shows that
Hypothesis~\ref{h3.1} is satisfied. Moreover, one obtains
\begin{align}
&g(z,x)=-\frac{s(z,x+\omega,x)}{2(\Delta(z)^2-1)^{1/2}},
\lb{3.44h} \\
&F_\infty (z,x)=-Cs(z,x+\omega,x), \quad
R_\infty (z)=C^2(\Delta(z)^2-1) \lb{3.44i}
\end{align}
for some constant $C\in\bbR.$ Hypothesis~\ref{h3.2}\,(i), (ii)
then follows from standard Floquet theory and
Hypothesis~\ref{h3.2}\,(iii), (iv) from iterating the Volterra
integral equations for $s(z,x,x_0)$ and $c(z,x,x_0),$
integrating by parts, and applying the Riemann-Lebesgue
lemma. Hence the Burchnall-Chaundy  theory for the
algebro-geometric case naturally extends to the  class of infinite
genus curves $\calK_\infty$ treated in this  section. Moreover,
every periodic $V\in C^1(\bbR)$ satisfies a stationary equation of
the type \eqref{3.44}.
\end{remark}
%%%%%%%%%%%%%%%%%%%%%%%%%%%%%%%%%%%%%%%%%%%%%%%%%%%%%%%%%%%%%%

For the time-dependent case one can closely follow the last
part of Section~\ref{s2} and hence we merely sketch the
corresponding steps. First we note that
\eqref{2.30}--\eqref{2.32} of course still apply and hence
it remains to consider the initial value problem with stationary
$\sKdV_\infty$ initial data. In complete analogy to
\eqref{2.37}--\eqref{2.40} one then introduces ($(x,t_r)\in\Omega$)
\begin{align}
&(1/2)F_{\infty,xxx}(0,x,t_r) -2V(x,t_r)F_{\infty,x}(0,x,t_r)
-V_x(x,t_r)F_\infty(0,x,t_r)=0, \lb{3.45} \\
&V_{t_r}(x,t_r)=-(1/2){\wti
F}_{r,xxx}(0,x,t_r)  +2V(x,t_r){\wti F}_{r,x}(0,x,t_r)
 +V_x(x,t_r){\wti F}_r (0,x,t_r). \lb{3.46}
\end{align}
In analogy to Section~\ref{s2}, \eqref{3.45} and \eqref{3.46} are
equivalent to
\begin{align}
\sKdV_\infty (V)&=0, \quad (x,t_r)\in\Omega,
\lb{3.47} \\
\wti \KdV_r (V)&=V_{t_r}-2{\ti f}_{r+1,x}=0,
\quad (x,t_r)\in\Omega. \lb{3.48}
\end{align}
Hence $V(x,t_r),$ for each fixed $t_r$ (with $(x,t_r)\in\Omega$)
satisfies the stationary $\sKdV_\infty$ equation, while
$V(x,t_r),$ $(x,t_r)\in\Omega,$ satisfies the time-dependent
$\KdV_r$  equation.

The analog of the time-dependent Baker-Akhiezer function
\eqref{2.32a} then is still given by
\begin{align}
&\psi(P,x,x_0,t_r,t_{0,r})=\exp\bigg(\int_{x_0}^x dx^\prime \,
\phi(P,x^\prime,t_r)+\int_{t_{0,r}}^{t_r}ds\,
({\wti F}_r (z,x_0,s)\phi(P,x_0,s) \no \\
&\hspace*{1cm} -(1/2){\wti F}_{r,x}  (z,x_0,s)) \bigg),
\quad P=(z,y)\in\calK_\infty,\,
(x,t_r), (x_0,t_{0,r})\in\Omega, \lb{3.49}
\end{align}
with $\phi(P,x,t_r)$ given as in \eqref{3.14}, \eqref{3.15}
(replacing $F_\infty(z,x)$  by $F_\infty(z,x,t_r)$ and
$H_\infty (z,x)$ by $H_\infty (z,x,t_r),$ respectively). As in
Section~\ref{s2}, $\psi(\cdot,x,x_0,t_r,t_{0,r})$ is meromorphic on
$\calK_\infty$ and
$(L(t_r)-z(P))\psi(P,\cdot,x_0,t_r,t_{0,r})=0.$ Next we
formulate the analogs of \eqref{2.32b}--\eqref{2.32d} in the
following Lemma.

%%%%%%%%%%%%%%%%%%%%%%%%%%%%%%%%%%%%%%%%%%%%%%%%%%%%%%%%%%%%%%%%%%
\begin{lemma} \lb{l3.7}
Assume Hypothesis~\ref{h3.1} and \ref{h3.2} (replacing
$\mu_j(x),$
$g(P,x),$ and  the open interval $\calI\subseteq\bbR$ by
$\mu_j(x,t_r),$
$g(P,x,t_r),$ and the open subset $\Omega\subseteq\bbR^2,$
respectively) supposing $\partial_x^k V\in C(\Omega),$
$k=0,\dots,2r+1,$ $V_{t_r}\in C(\Omega).$ In addition, let
$P=(z,y)\in\calK_\infty,$ $(x,t_r), (x_0,t_{0,r})\in\Omega.$ Then
Lemmas~\ref{l3.3} and \ref{l3.4} apply to
$\phi(P,x,t_r),$ $\psi(P,x,x_0,t_r,t_{0,r}),$ $g(P,x,t_r),$
and $V(x,t_r).$ Moreover, one has
\begin{align}
&\phi_{t_r}(P,x,t_r)=\partial_x({\wti F}_r (z,x,t_r)
\phi(P,x,t_r)-(1/2){\wti F}_{r,x} (z,x,t_r)), \lb{3.50} \\
&\psi_{t_r}(P,x,x_0,t_r,t_{0,r})=(\wti
P_{2r+1}(t_r)\psi)(P,x,x_0,t_r,t_{0,r}) \lb{3.51} \\
&=\ul{\wti F}_r (z,x,t_r)\psi_x(P,x,x_0,t_r,t_{0,r})
 -(1/2){\wti F}_{r,x}
(z,x,t_r)\psi(P,x,x_0,t_r,t_{0,r}), \no \\
&F_{\infty,t_r} (z,x,t_r)={\wti F}_r (z,x,t_r) F_{\infty,x}
(z,x,t_r)  -{\wti F}_{r,x} (z,x,t_r) F_\infty (z,x,t_r).
\lb{3.52}
\end{align}
\end{lemma}
%%%%%%%%%%%%%%%%%%%%%%%%%%%%%%%%%%%%%%%%%%%%%%%%%%%%%%%%%%%%%%%%%%
%%%%%%%%%%%%%%%%%%%%%%%%%%%%%%%%%%%%%%%%%%%%%%%%%%%%%%%%%%%%%%%%%%
\begin{proof}
We only need to prove \eqref{3.50}--\eqref{3.52}. Starting from
\begin{align}
V_{t_r}(x,t_r)&=-(1/2){\wti
F}_{r,xxx}(z,x,t_r)  +2(V(x,t_r)-z){\wti F}_{r,x}(z,x,t_r)
\no \\
&\quad \, +V_x(x,t_r){\wti F}_r (z,x,t_r), \lb{3.53}
\end{align}
which is equivalent to \eqref{3.45} due to the recursion
\eqref{2.2}, one computes from \eqref{3.17},
\begin{align}
&\partial_{t_r}(\phi_x+\phi^2)=\phi_{t_r,x}+2\phi\phi_{t_r}=V_{t_r}
\no \\
&=({\wti F}_r\phi-(1/2){\wti F}_r)_{xx}+2\phi({\wti
F}_r\phi-(1/2){\wti F}_r)_x. \lb{3.54}
\end{align}
Thus,
\begin{equation}
(\partial_x +2\phi)(\phi_{t_r}-({\wti F}_r\phi-(1/2)
{\wti F}_r)_x)=0 \lb{3.55}
\end{equation}
and hence
\begin{equation}
\phi_{t_r}=({\wti F}_r\phi-(1/2)
{\wti F}_r)_x +C\exp \bigg(-2\int^x dx^\prime\,\phi\bigg),
\lb{3.56}
\end{equation}
where $C$ is independent of $x$ (but may depend on $P$ and $t_r$).
Next, combining \eqref{3.21} and the Riccati equation \eqref{3.17},
one observes that $\phi_x$ has an asymptotic expansion of the type
\begin{equation}
\phi_x(P,x,t_r)=o(z(P)) \lb{3.56a}
\end{equation}
uniformly with respect to $(x,t_r)$ as long as $(x,t_r)$ varies in
compact subsets of $\Omega.$ Inserting \eqref{3.56a} into
\eqref{3.56} and integrating with respect to $t_r$ then
contradicts the asymptotic expansion \eqref{3.21} for $\phi$
unless $C=0.$ This proves \eqref{3.50}. \eqref{3.51}
is then obvious from \eqref{3.49} and \eqref{3.50}. Using
\eqref{3.20} and \eqref{3.50}, one obtains
\begin{align}
&\phi_{t_r}(P)-\phi_{t_r}(P^*)=2y(P)F_\infty^{-2} F_{\infty,t_r}
\no \\
&=\partial_x({\wti F}_r(\phi(P)-\phi(P^*)))=2y(P)F_\infty^{-2}
({\wti F}_rF_{\infty,x}-{\wti F}_{r,x}F_\infty) \lb{3.57}
\end{align}
and hence \eqref{3.52}.
\end{proof}
%%%%%%%%%%%%%%%%%%%%%%%%%%%%%%%%%%%%%%%%%%%%%%%%%%%%%%%%%%%%%%%%

%%%%%%%%%%%%%%%%%%%%%%%%%%%%%%%%%%%%%%%%%%%%%%%%%%%%%%%%%%%%%%
\begin{remark} \lb{r3.8}
As in Remark~\ref{r2.1} we emphasize that the
isospectral deformation problem \eqref{3.47}, \eqref{3.48} (or
equivalently, \eqref{3.45}, \eqref{3.46}) is not necessarily
well-posed. Unique solvability of \eqref{3.47}, \eqref{3.48} can
again be reduced to unique solvability of the corresponding
first-order system of Dubrovin equations. In the present
infinite genus context this system is of the form
\begin{align}
\mu_{j,x}(x,t_r)&=
\frac{-2y(\hat \mu_j(x,t_r))}
{\prod_{\ell\in\bbN\backslash\{j\}} (\mu_\ell(x,t_r)-
\mu_j(x,t_r))/(E_{2\ell-1}E_{2\ell})^{1/2}},\lb{3.58}\\
\mu_{j,t_r}(x,t_r)&=
{\wti F}_r(\mu_j(x,t_r),x,t_r)\mu_{j,x}(x,t_r)\label{3.59}\\
&=\frac{-2y(\hat \mu_j(x,t_r))}
{\prod_{\ell\in\bbN\backslash\{j\}} (\mu_\ell(x,t_r)-
\mu_j(x,t_r))/(E_{2\ell-1}E_{2\ell})^{1/2}}{\wti F}_r(\mu_j(x,t_r),x,t_r) \no
\end{align}
for $j\in\bbN,$ with initial data for
\eqref{3.58}, \eqref{3.59} on $\calK_\infty$ given by
\begin{equation}
\{\hat \mu_1 (x_0,t_{0,r})\}_{j\in\bbN}\subset\calK_\infty.
\lb{3.60}
\end{equation}
Here $\hat \mu_j (x,t_r))\in\calK_\infty$ is defined as
in \eqref{3.15a} (replacing $x$ by $x,t_r$)
\begin{equation}
\hat \mu_j(x,t_r)=(\mu_j(x,t_r),
-(1/2)F_{\infty,x}(\mu_j(x,t_r),x,t_r)),
\quad j\in\bbN, \lb{3.61}
\end{equation}
and \eqref{3.58} is then seen to be an
immediate consequence of the analog of \eqref{3.8}, that is,
\begin{equation}
F_\infty (z,x,t_r)=\prod_{j\in\bbN}
((\mu_j(x,t_r)-z)/(E_{2j-1}E_{2j})^{1/2}), \lb{3.61b}
\end{equation}
and of \eqref{3.61}, while
\eqref{3.59} follows from \eqref{3.52}, taking $z=\mu_j(x,t_r)$
in either case. As long as the curve $\calK_\infty$ is
nonsingular (i.e.,
$E_m\neq E_{m^\prime}$ for $m\neq m^\prime$) and none of the
$\mu_j(x,t_r)$ collide (i.e., for $j\neq j^\prime,$
$\mu_j(x,t_r)\neq\mu_{j^\prime}(x,t_r)$ for $(x,t_r)\in\Omega,$
$\Omega\in\bbC^2$ open), the Dubrovin system is
uniquely solvable on $\calK_\infty$ and $V(x,t_r),$
$(x,t_r)\in\Omega,$ is then recovered from the time-dependent trace
formula
\begin{equation}
V(x,t_r)=E_0+\sum_{j\in\bbN}(E_{2j-1}+E_{2j}-2\mu_j(x,t_r)).
\lb{3.62}
\end{equation}
The corresponding $F_\infty(z,x,t_r),$
${\wti F}_r(z,x,t_r),$ $(x,t_r)\in\Omega,$ then satisfy the
isospectral deformation  equations \eqref{3.47}, \eqref{3.48}.
Such ideas, in the case of nonsingular curves
$\calK_\infty$ with real branch points
$\{E_m\}_{m\in\bbN}\subset\bbR,$ $E_m\neq E_{m^\prime}$ for
$m\neq m^\prime$ have been applied by Trubowitz \cite{Tr77} in the
stationary periodic case, by Kotani and Krishna \cite{KK88} in some
almost periodic stationary cases, and in great
generality by Sodin and Yuditskii \cite{SY95} (avoiding
Dubrovin equations and directly working with the corresponding
infinite dimensional Jacobi inversion problem) in the case $r=0.$
The general case of the KdV  hierarchy is discussed by Egorova
\cite{Eg94}, Levitan
\cite{Le82}, \cite{Le83}, \cite{Le84}, \cite{Le85}, \cite[Ch.~11,
12]{Le87}, and in the periodic case by
Marchenko \cite[Sect.~4.3]{Ma86}.
\end{remark}
%%%%%%%%%%%%%%%%%%%%%%%%%%%%%%%%%%%%%%%%%%%%%%%%%%%%%%%%%%%%%%

The isospectral deformation problem \eqref{3.47}, \eqref{3.48}
clearly deserves further study. For instance, suppose $V(x,0)$ is
real-analytic (or elliptic) and of infinite genus type satisfying
\eqref{3.47} (resp. \eqref{3.45}). A natural question to ask is
whether the solution $V(x,t_r)$ of \eqref{3.47}, \eqref{3.48}
is real-analytic (or elliptic) with respect to $x$ for each
$t_r\in\bbR.$ Similarly, on the stationary level, one might ask
whether the isospectral torus of a real-analytic (or elliptic) and
infinite genus type potential $V(x)$ satisfying
\eqref{3.44} consists of only real-analytic (or elliptic)
potentials. (In the real-valued finite genus case and in the case
of real-valued (infinite genus) periodic potentials
the answer is well-known to  be affirmative, cf.~\cite{BF84},
\cite{Tr77}.)

We did not consider the infinite genus limit $n\to\infty$ of the
theta function representation (i.e., the Its-Matveev formula)
\eqref{2.41} of $V(x,t_r)$ in this paper but plan to return to this
topic elsewhere.

%%%%%%%%%%%%%%%%%%%%%%%%%%%%%%%%%%%%%%%%%%%%%
\section{Darboux-type transformations} \lb{s4}
%%%%%%%%%%%%%%%%%%%%%%%%%%%%%%%%%%%%%%%%%%%%%

In this section we study the effects of Darboux transformations of
the KdV  hierarchy on the transcendental hyperelliptic
curves introduced in Section~\ref{s3}.

Since the the algebro-geometric (finite genus) case has recently
been treated in detail in \cite{GH98a}, we exclusively focus on the
infinite genus case in this section. Moreover, since time plays no
active role in the following arguments, we shall suppress
$t_r$ in what follows and without loss of generality work within
the stationary formalism.

Assuming Hypothesis~\ref{h3.2}, we start by introducing Darboux
transformations in connection with  the differential expression
$L=-d^2/dx^2+V$ and the curve $\calK_\infty$ defined in
\eqref{3.9}. Define (for $P\in \calK_\infty$)
\begin{equation}
\psi(P,x,x_{0},\sigma)= \begin{cases}
\tfrac{1}{2}(1+\sigma)\psi(P,x,x_{0})
+\tfrac{1}{2}(1-\sigma)\psi(P^*,x,x_{0}),
& \sigma\in \bbC,  \\
\psi(P,x,x_{0})-\psi(P^*,x,x_{0}), &\sigma=\infty, \end{cases}
\lb{4.35}
\end{equation}
pick $Q_{0}=(z_0,y_0)\in \calK_\infty$, and
introduce the
differential expressions
\begin{equation}
A_{\sigma}(Q_{0})=\f{d}{dx}+\phi(Q_{0},x,\sigma), \quad
A_{\sigma}^+(Q_{0})=-\f{d}{dx}+\phi(Q_{0},x,\sigma), \quad
\sigma\in\bbCinf, \lb{4.36}
\end{equation}
where
\begin{equation}
\phi(P,x,\sigma)=\psi_{x}(P,x,x_{0},\sigma)/\psi(P,x,x_{0},\sigma),
\quad P\in \calK_\infty, \,
\sigma\in\bbCinf. \lb{4.37}
\end{equation}
One verifies (cf.\ \eqref{3.17})
\begin{equation}
L=A_{\sigma}(Q_{0})A_{\sigma}^+(Q_{0})+z_{0}=
-\f{d^2}{dx^2}+V,
\lb{4.38}
\end{equation}
with
\begin{equation}
V(x)=\phi(Q_{0},x,\sigma)^2+\phi_{x}(Q_{0},x,\sigma)+z_{0},
\lb{4.39}
\end{equation}
independent of the choice of $\sigma\in\bbCinf$.
Interchanging the order of the differential expressions
$A_{\sigma}(Q_{0})$ and $A_{\sigma}^+(Q_{0})$ in
\eqref{4.38} then
yields
\begin{equation}
\widehat L_{\sigma}(Q_{0})=A_{\sigma}^+(Q_{0})
A_{\sigma}(Q_{0})+z_{0}
=-\f{d^2}{dx^2}+\widehat V_{\sigma}(x,Q_{0}), \lb{4.40}
\end{equation}
with
\begin{align}
\widehat V_{\sigma}(x,Q_{0})&=\phi(Q_{0},x,\sigma)^2-
\phi_{x}(Q_{0},x,\sigma)+z_{0}\no \\
&=V(x)-2(\ln(\psi(Q_{0},x,x_{0},\sigma)))_{xx}, \quad
\sigma\in\bbCinf.\lb{4.41}
\end{align}
The transformation
\begin{equation}
V(x)\mapsto \widehat V_{\sigma}(x,Q_{0}), \quad
Q_{0}\in \calK_\infty, \,
\sigma\in\bbCinf \lb{4.42}
\end{equation}
is usually called the Darboux transformation (also Crum-Darboux
transformation or single commutation method) and goes back to at
least Jacobi  \cite{Ja46} and Darboux \cite{Da82}. While we only
aim at its properties from an algebraic point of view, its analytic
properties in connection with spectral deformations (isospectral
and  non-isospectral
ones) have received enormous attention in the context of spectral
theory (especially, regarding the insertion of eigenvalues into
spectral gaps), inverse spectral theory, and  B\"acklund
tranformations for the (time-dependent) KdV hierarchy.  A complete
bibliography in this context being impossible, we just refer to
\cite{Cr55}, \cite{De78}, \cite{DT79},
\cite[Ch.~4]{EK82a}, \cite{EK82},
\cite{EF85}, \cite{FM76}, \cite{Ge93}, \cite{GSS91},
\cite{GST96}, \cite{GS95}, \cite{GW93}, \cite{Mc85}, \cite{Mc86},
\cite{Mc87}, \cite{Sc78}, and the
extensive literature therein. From a historical point of view it is
very interesting to note that Drach \cite{Dr18}, \cite{Dr19},
\cite{Dr19a} in his 1919 studies of Darboux transformations
 (being a student of Darboux) not only introduced a set of
nonlinear differential equations for $V,$ which today can be
identified with the stationary KdV hierarchy, but also studied the
effect of Darboux transformations on the underlying hyperelliptic
curve. As a consequence, he seems to have been the first to
explicitly establish the connection between integrable systems and
spectral theory. For recent treatments of this connection see, for
instance, \cite[Ch.~3]{BBEIM94} and \cite{GH00}.

Assuming that $\psi\in\ker(L-z)$,
\begin{equation}
L\psi(z)=z\psi(z), \lb{4.47}
\end{equation}
one infers
$A_{\sigma}^+(Q_{0})\psi(z)\in\ker(\widehat
L_{\sigma}(Q_{0})-z)$,
\begin{equation}
\widehat L_{\sigma}(Q_{0})(A_{\sigma}^+(Q_{0})\psi(z))
=zA_{\sigma}^+(Q_{0})\psi(z), \lb{4.48}
\end{equation}
and
\begin{align}
&W(A_{\sigma}^+(Q_{0})\psi_{1}(z),
A_{\sigma}^+(Q_{0})\psi_{2}(z))
=(z-z_{0})W(\psi_{1}(z),\psi_{2}(z)), \lb{4.49} \\
&\hspace*{5.25cm}\psi_{1}(z),\psi_{2}(z)\in\ker(L-z). \no
\end{align}
Since
\begin{equation}
(A_{\sigma}^+(Q_{0})\psi(P,\dott,x_{0}))(x)
=\big(\phi(Q_{0},x,\sigma)-\phi(P,x)\big)\psi(P,x,x_{0})
 \lb{4.50}
\end{equation}
for $P\in\calK_\infty\backslash\{Q_{0}\},$ we define
\begin{align}
\hat\psi_{\sigma}(P,x,x_{0},Q_{0}) &=
(A_{\sigma}^+(Q_{0})\psi(P,\dott,x_{0}))(x) \lb{4.51} \\
&=\big(\phi(Q_{0},x,\sigma)-\phi(P,x)\big)
\psi(P,x,x_{0}),  \quad  P\in\calK_\infty\backslash\{Q_{0}\}, \,
\sigma\in\bbCinf. \no
\end{align}
Then
\begin{equation}
(\widehat L_{\sigma}(Q_{0})\hat
\psi_{\sigma}(P,\dott,x_{0},Q_{0}))(x)
=z\hat\psi_{\sigma}(P,x,x_{0},Q_{0}) \lb{4.52}
\end{equation}
for $P=(z,y)\in\calK_\infty\backslash\{Q_{0}\},$ and we define in
analogy to \eqref{3.7} the diagonal Green's  function
$\hat g_{\sigma}(P,x,Q_{0})$ of $\widehat
L_{\sigma}(Q_{0})$ by
\begin{equation}
\hat g_{\sigma}(P,x,Q_{0})=
\f{\hat\psi_{\sigma}(P,x,x_{0},Q_{0})
\hat\psi_{\sigma}(P^{*},x,x_{0},Q_{0})}
{W(\hat\psi_{\sigma}(P,\dott,x_{0},Q_{0}),
\hat\psi_{\sigma}(P^{*},\dott,x_{0},Q_{0}))},
\quad P=(z,y)\in\calK_\infty\backslash\{Q_{0}\}. \lb{4.53}
\end{equation}

%%%%%%%%%%%%%% lemma %%%%%%%%%%%%%%%%%%%%%%%%%%%%%%%%%%%%%%
\begin{lemma}\lb{l4.2} Assume $\sKdV_\infty (V)=0$ and let
$Q_{0}=(z_{0},y_{0})\in\calK_\infty$,
$P=(z,y)\in\calK_\infty\backslash\{Q_{0}\}$,
$\sigma\in\bbCinf$. Then the diagonal Green's function
$\hat g_{\sigma}(P,x,Q_{0})$ in \eqref{4.53} explicitly
reads
\begin{align}
\hat g_{\sigma}(P,x,Q_{0})
&=\f{(\phi(P,x)-\phi(Q_{0},x,\sigma))(\phi(P^{*},x)-
\phi(Q_{0},x,\sigma))
F_\infty(z,x)}{2(z-z_{0})y(P)} \lb{4.54b}\\
&=\f{H_\infty(z,x)+\phi(Q_{0},x,\sigma)^2F_\infty(z,x)
-\phi(Q_{0},x,\sigma)F_{\infty,x}(z,x)}
{2(z-z_{0})y(P)} \lb{4.54a} \\
&=\f{\widehat F_{\sigma,\infty}(z,x,Q_0)}{2\hat y(P)},
\lb{4.54c}
\end{align}
where $\hat y(P)$ denotes the meromorphic solution on
$\widehat\calK_{\sigma,\infty}(Q_{0})$ obtained
upon solving
$y^2=\widehat R_{\sigma,\infty}(z,Q_0)$, $P=(z,y)$ for some
entire function $\widehat R_{\sigma,\infty}(z,Q_0)$ of the
type \eqref{3.9} and
$\widehat F_{\sigma,\infty}(z,x,Q_0)$ denotes an entire
function with respect to $z$ of the type \eqref{3.8}. In
particular, the Darboux  transformation \eqref{4.42},
$V(x)\mapsto\widehat V_{\sigma}(x,Q_{0})$  maps the class
of $\KdV_\infty$ potentials into itself.
\end{lemma}
%%%%%%%%%%%%%%%%%%%%%%%%%%%%%%%%%%%%%%%%%%%%%%%%%%%%%%%%%%%%%%%
\begin{proof}
\eqref{4.54b} and \eqref{4.54a} follow upon use of
$\phi(P,x)=\psi_{x}(P,x,x_{0})/\psi(P,x,x_{0})$, \eqref{3.18},
\eqref{3.19},
\eqref{3.23}--\eqref{3.26}, and \eqref{4.49}.  Since the
numerator in
\eqref{4.54a} is entire in $z$, and
\begin{equation}
\hat g_{\sigma}(P,x,Q_{0})=(i/2)\zeta (P)+
O(\zeta (P)^2) \text{  as $P\to P_{\infty}$} \lb{4.55}
\end{equation}
again by \eqref{4.54a}, one concludes \eqref{4.54c}.
By inspection, $\widehat F_{\sigma,\infty}(z,x,Q_0)$
satisfies equation \eqref{3.12} (and hence \eqref{3.44})
with $V(x)$ replaced  by
$\widehat V_{\sigma}(x,Q_{0})$ and  $R_\infty(z)$
by $\widehat R_{\sigma,\infty}(z,Q_0)$. Consequently,
$V(x)$ being a $\KdV_\infty$ potential implies that
$\widehat  V_{\sigma}(x,Q_{0})$ is one as well.
\end{proof}
%%%%%%%% end of lemma %%%%%%%%%%%%%%%%%%%%%%%%%%%%%%%%%%%%%%%%

The following theorem, the principal result of this section, will
clarify the relation between $\calK_\infty$ and $\widehat
\calK_{\sigma,\infty}(Q_0)$ depending on $(Q_{0},\sigma).$

%%%%%%%% theorem %%%%%%%%%%%%%%%%%%%%%%%%%%%%%%%%%%%%%%%%%%%%%
\begin{theorem}\lb{t4.3}
Suppose $\sKdV_\infty (V)=0$ and let
$Q_{0}=(z_{0},y_{0})\in\calK_\infty$ and
$\sigma\in\bbCinf.$ Then the transcendental hyperelliptic curve
$\widehat\calK_{\sigma,\infty}(Q_{0})$
associated with $\widehat V_{\sigma}(x,Q_{0})$ is of the type
\begin{equation}
\widehat\calK_{\sigma,\infty}(Q_0)\colon
\widehat \calF_{\sigma,\infty}(z,y,Q_{0})=y^2-
\widehat R_{\sigma,\infty}(z,Q_{0})=0, \lb{4.56a}
\end{equation}
with
\begin{align}
\widehat R_{\sigma,\infty}(z,Q_{0})
& =\begin{cases}
(z-z_{0})^2 R_\infty (z), & \sigma\in\bbCinf\backslash\{-1,1\}
\text{ and } y_{0}\neq 0, \\
(z-z_{0})^2 R_\infty (z), & \sigma=\infty \text{ and }
y_{0}= 0, \\
R_\infty (z), & \sigma\in\{-1,1\} \text{ and } y_{0}\neq 0, \\
R_\infty (z), & \sigma\in\bbC, \, y_{0}=0, \text{ and }
R_{\infty,z}(z_{0})\neq 0, \\
(z-z_{0})^{-2} R_\infty (z), & \sigma\in\bbC, \, y_{0}=0,
\text{ and } R_{\infty,z}(z_{0})=0,
\end{cases} \lb{4.57}
\end{align}
and $R_\infty (z) =\prod_{m=0}^{2n}(1-(z/E_m)).$
\end{theorem}
%%%%%%%%%%%%%%%%%%%%%%%%%%%%%%%%%%%%%%%%%%%%%
\begin{proof}  Our starting point will be \eqref{4.54b} and
a careful
case distinction taking into account whether or not $Q_{0}$ is a
branch point, and distinguishing the cases
$\sigma\in\bbC\backslash\{-1,1\}$, $\sigma\in\{-1,1\}$, and
$\sigma=\infty$.

Case (i).  $\sigma\in\bbCinf\backslash\{-1,1\}$ and
$y_{0}\neq 0$:  One
computes from \eqref{4.35} and \eqref{4.37},
\begin{equation}
\phi(Q_{0},x,\sigma)=\begin{cases}
\f{(1+\sigma)\psi_{x}(Q_{0},x,x_{0})+
(1-\sigma)\psi_{x}(Q_{0}^{*},x,x_{0})}
{(1+\sigma)\psi(Q_{0},x,x_{0})+(1-\sigma)\psi(Q_{0}^{*},x,x_{0})},
& \sigma\in\bbC\backslash\{-1,1\}, \\[2mm]
\f{\psi_{x}(Q_{0},x,x_{0})-\psi_{x}(Q_{0}^{*},x,x_{0})}
{\psi(Q_{0},x,x_{0})-\psi(Q_{0}^{*},x,x_{0})},
& \sigma=\infty,
\end{cases} \lb{4.58}
\end{equation}
and upon comparison with $\phi(Q_{0},x)\neq\phi(Q_{0}^{*},x)$,
\begin{equation}
\phi(Q_{0},x)=\f{\psi_{x}(Q_{0},x,x_{0})}{\psi(Q_{0},x,x_{0})},
\quad
\phi(Q_{0}^{*},x)=\f{\psi_{x}(Q_{0}^{*},x,x_{0})}
{\psi(Q_{0}^{*},x,x_{0})},\lb{4.59}
\end{equation}
one concludes that no cancellations can occur in \eqref{4.54b},
proving the first statement in \eqref{4.57}.

Case (ii).  $\sigma=\infty$ and $y_{0}=0$: Combining
\eqref{3.14},
\eqref{3.16}, \eqref{3.20}, and \eqref{4.35} one computes
\begin{align}
&\phi(Q_{0},x,\infty)=\lim_{P\to Q_{0}}\phi(P,x,\infty) \no\\
&=\lim_{P\to Q_{0}}\Bigg(
\f{\phi(P,x)\exp\big(\int_{x_{0}}^xdx'\,\phi(P,x')\big)-
\phi(P^{*},x)\exp\big(\int_{x_{0}}^xdx'\,\phi(P^{*},x')\big)}
{\exp\big(\int_{x_{0}}^xdx'\,\phi(P,x')\big)-
\exp\big(\int_{x_{0}}^xdx'\,\phi(P^{*},x')\big)}\Bigg)	\no \\
&=\phi(Q_{0},x)\no \\
&+\lim_{P\to Q_{0}}\Bigg(\f{\phi(P,x)-\phi(P^{*},x)}
{\exp\big(\int_{x_{0}}^xdx'\,\phi(P,x')\big)-
\exp\big(\int_{x_{0}}^xdx'\,\phi(P^{*},x')\big)}\times \no \\
& \hspace*{5cm} \times \exp\bigg(\int_{x_{0}}^x
dx'\,\phi(P^{*},x')\bigg)\Bigg)
\no
\\ &=\phi(Q_{0},x)+\exp\bigg(\int_{x_{0}}^xdx'\,
\phi(Q_{0},x')\bigg)\times \no \\
&\qquad\qquad\times\lim_{P\to Q_{0}}\Bigg(\f{-2y(P)/F_\infty(z,x)}
{\exp\big(-y(P)\int_{x_{0}}^x\f{dx'}{F_\infty(z,x')}\big)
-\exp\big(y(P)\int_{x_{0}}^x\f{dx'}{F_\infty(z,x')}\big)}\times
\no \\
&\hspace*{6.15cm} \times\exp\bigg(-\f12\int_{x_{0}}^xdx'\,
\f{F_{\infty,x}(z,x')}{F_\infty(z,x')}\bigg)\Bigg)	\no \\
&=\phi(Q_{0},x)\no \\
&\qquad+\psi(Q_{0},x,x_{0})\f{1}{F_\infty(z_{0},x)
\psi(Q_{0},x,x_{0})}
\lim_{P\to Q_{0}}\bigg(\f{2y(P)}{2y(P)\int_{x_{0}}^x\f{dx'}
{F_\infty(z,x')}
+O(y(P)^2)}\bigg)	\no \\
&=\phi(Q_{0},x)+\bigg(F_\infty(z_{0},x)\int_{x_{0}}^x\f{dx'}
{F_\infty(z,x')}\bigg)^{-1},
\quad x\in\bbC\backslash\{x_{0}\},  \lb{4.60}
\end{align}
using $\lim_{P\to Q_{0}} y(P)=y(Q_{0})=y_{0}=0$.   From
\begin{equation}
\phi(Q_{0},x)=\f12 \f{F_{\infty,x}(z_{0},x)}{F_\infty(z_{0},x)}
\lb{4.61}
\end{equation}
one concludes again that no cancellations can occur in
\eqref{4.54b}.
Thus, the second statement in \eqref{4.57} holds.

The remainder of the proof requires a more refined argument,
the basis of which will be derived next. Writing
\begin{equation}
y(P)^2=R_\infty(z)\underset{z\to z_0}{=} y_{0}^2+
\hat y_{1}(z-z_0)+\hat y_{2}(z-z_0)^2+O((z-z_0)^3), \lb{4.62}
\end{equation}
a comparison of the powers $(z-z_0)^0$ and $(z-z_0)^1$
in \eqref{3.8}
yields
\begin{equation}
2F_{\infty,xx}(z_0,x)F_\infty(z_0,x)-F_{\infty,x}(z_0,x)^2
-4VF_\infty(z_0,x)^2+4y_{0}^2=0	\lb{4.63}
\end{equation}
and
\begin{align}
&\dot F_{\infty,xx}(z_0,x)F_\infty(z_0,x)+F_{\infty,xx}(z_0,x)\dot
F_\infty(z_0,x)-F_{\infty,x}(z_0,x)\dot F_{\infty,x}(z_0,x) \no \\
&-4VF_\infty(z_0,x)\dot F_\infty(z_0,x)
+2F_\infty(z_0,x)^2+2\hat y_{1}=0.	\lb{4.64}
\end{align}
Inserting \eqref{4.63} into \eqref{4.64}, a little algebra
proves the
basic identity
\begin{align}
&F_\infty(z_0,x)^2(\dot F_\infty(z_0,x)/F_\infty(z_0,x))_{xx}
+F_{\infty,x}(z_0,x)F_\infty(z_0,x)(\dot
F_\infty(z_0,x)/F_\infty(z_0,x))_{x} \no \\
&	+2F_\infty(z_0,x)^2
-4y_{0}^2(\dot F_\infty(z_0,x)/F_\infty(z_0,x))
+2\hat y_{1}=0.	\lb{4.65}
\end{align}

Case (iii). $\sigma\in\{-1,1\}$ and $y_{0}\neq 0$: Then
\eqref{4.35}
yields
\begin{equation}
	\phi(Q_{0},x,1)=\phi(Q_{0},x), \quad \phi(Q_{0},x,-1)=
\phi(Q_{0}^{*},x),
		\lb{4.66}
\end{equation}
with $\phi(Q_{0},x)\neq\phi(Q_{0}^{*},x)$ since $y_{0}\neq 0$. In
this case there is a cancellation in \eqref{4.54b}.  For instance,
choosing $\sigma=1$ one computes from \eqref{3.8} and
\eqref{3.14},
\begin{align}
\phi(P,x)-\phi(Q_{0},x,&1)=\phi(P,x)-\phi(Q_{0},x) \no \\
&\underset{P\to
Q_{0}}{=}\f{-(y(P)-y_{0})}{F_\infty(z_0,x)}
+y_{0}\f{\dot F_\infty(z_0,x)}{F_\infty(z_0,x)^2}(z-z_0)
\no \\
&\hspace*{9.7mm} +\f12
\bigg(\f{\dot F_\infty(z_0,x)}{F_\infty(z_0,x)}\bigg)_{x}(z-z_0)
+O((z-z_0)^2) \no \\
&\underset{P\to Q_{0}}{=}c_{1}(x)(z-z_0)+O((z-z_0)^2) \lb{4.67}
\end{align}
since
\begin{equation}
y(P)-y_{0}\underset{P\to Q_{0}}{=}y_{1}(z-z_0)+O((z-z_0)^2), \quad
\hat  y_{1}=2y_{0}y_{1}.	\lb{4.68}
\end{equation}
It remains to show that $c_{1}(x)$ does not vanish identically in
$x\in\bbC$.  Arguing by contradiction we assume
\begin{align}
0&=c_{1}(x) \lb{4.69}	 \\
&=\f{-y_{1}}{F_\infty(z_0,x)}+
	\f{y_{0}}{F_\infty(z_0,x)}\f{\dot F_\infty(z_0,x)}
{F_\infty(z_0,x)}
	+\f12\bigg(\f{\dot F_\infty(z_0,x)}{F_\infty(z_0,x)}
\bigg)_{x}, \quad x\in\bbC. \no
\end{align}
Differentiating \eqref{4.69} with respect to $x$ and inserting
the ensuing expression for $(\dot F_\infty(z_0,x)/F_\infty(z_0,x))_{xx}$ and
the one for
$(\dot F_\infty(z_0,x)/F_\infty(z_0,x))_{x}$ from \eqref{4.69} into
\eqref{4.65} then
results in the contradiction
\begin{equation}
	0=2F_\infty(z_0,x)^2, \quad x\in\bbC.\lb{4.70}
\end{equation}
Moreover, since
\begin{equation}
\phi(P^{*},x)-\phi(Q_{0},x,x,1)=\phi(P^{*},x)-\phi(Q_{0},x)
\underset{P\to
Q_{0}}{=}\f{2y_{0}}{F_\infty(z_0,x)}+O(z-z_0),\lb{4.71}
\end{equation}
one concludes that precisely one factor of $z-z_0$ cancels in
\eqref{4.54b}.  Hence, the third relation in \eqref{4.57} holds.
The case $\sigma=-1$ in treated  analogously.

Case (iv). $\sigma\in\bbC$, $y_0=0$, and
$R_{\infty,z}(z_0)\neq 0$:   Taking into account that
$\phi(Q_{0},x,\sigma)=\phi(Q_{0},x)$  (using
\eqref{4.37} and $Q_0=Q_0^*$) is
independent of $\sigma\in\bbC$, \eqref{3.8} and \eqref{3.14}
yield
\begin{equation}
	\big(\phi(P,x)-\phi(Q_{0},x) \big) \big(\phi(P^{*},x)-
\phi(Q_{0},x)   \big)
\underset{P\to Q_{0}}{=}\f{-y_{1}^2
(z-z_0)}{F_\infty(z_0,x)^2}+O((z-z_0)^2)
\lb{4.72}
\end{equation}
since
\begin{equation}
	y(P)\underset{P\to Q_{0}}{=}y_{1}(z-z_0)^{1/2}+O((z-z_0)^{3/2}).
		\lb{4.73}
\end{equation}
Thus we infer again that precisely one factor of $z-z_0$ cancels in
\eqref{4.54b}.  Hence, the fourth relation in \eqref{4.57} is proved.

Case (v). $\sigma\in\bbC$, $y_{0}=\hat y_{1}=0$, and $\hat
y_{2}\neq 0$ (cf.\ \eqref{4.62}):  One calculates as in
\eqref{4.72},
\begin{align}
&\big(\phi(P,x)-\phi(Q_{0},x) \big) \big(\phi(P^{*},x)-
\phi(Q_{0},x)
\big) \no \\
& \quad\underset{P\to Q_{0}}{=} \bigg(\f{-y_{1}^2}
{F_\infty(z_0,x)^2}
+\f14 \bigg(\bigg(\f{\dot F_\infty(z_0,x)}
{F_\infty(z_0,x)}\bigg)_{x} \bigg)^2
\bigg)(z-z_0)^2+O((z-z_0)^3) \no \\
& \quad\underset{P\to Q_{0}}{=}c_{2}(x)(z-z_0)^2+O((z-z_0)^3)
\lb{4.74}
\end{align}
since
\begin{equation}
	y(P)\underset{P\to Q_{0}}{=}y_{1}(z-z_0)+O((z-z_0)^{2}).
		\lb{4.75}
\end{equation}
Next we show that $c_{2}(x)$ does not vanish identically in
$x\in\bbC$.  Arguing again by contradiction we suppose that
\begin{equation}
0=c_{2}(x)=\f{-y_{1}^2}{F_\infty(z_0,x)^2}
+\f{1}{4} \bigg(\bigg(\f{\dot F_\infty(z_0,x)}
{F_\infty(z_0,x)}\bigg)_{x} \bigg)^2,
\quad x\in\bbC.\lb{4.76}
\end{equation}
Thus
\begin{equation}
\bigg(\f{\dot F_\infty(z_0,x)}{F_\infty(z_0,x)}\bigg)_{x}
=\f{C}{F_\infty(z_0,x)}\lb{4.77}
\end{equation}
for some constant $C\in\bbC$.  Insertion of \eqref{4.77} and its
$x$-derivative into \eqref{4.65} then again yields the
contradiction
\begin{equation}
	0=2F_\infty(z_0,x)^2, \quad x\in\bbC.\lb{4.78}
\end{equation}
Hence the last relation in \eqref{4.57} holds in this case.

Case (vi). $\sigma\in\bbC$, $y_{0}=\hat y_{1}=\hat y_{2}=0$
(cf.\ \eqref{4.62}):  As in \eqref{4.74} one obtains
\begin{align}
\big(\phi(P,x)-\phi(Q_{0},x) \big) \big(\phi(P^{*},x)-
\phi(Q_{0},x) \big)&
\underset{P\to Q_{0}}{=}\f{1}{4} \bigg(\bigg(
\f{\dot F_\infty(z_0,x)}
{F_\infty(z_0,x)}\bigg)_{x}
\bigg)^2 (z-z_0)^2 \no \\
&\hspace*{9.9mm} +O((z-z_0)^3) \lb{4.79}
\end{align}
since
\begin{equation}
	y(P)\underset{P\to Q_{0}}{=}O((z-z_0)^{3/2}). \lb{4.80}
\end{equation}
The remainder of the proof of case (vi) is now just a special
case of case (v) (with $y_{1}=C=0$).
\end{proof}
%%%%%%%%% end of theorem %%%%%%%%%%%%%%%%%%%%%%%%%%%%%%%%%%%%%%

These results show, in particular, that Darboux
transformations do
not change the local structure of the original curve
$y^2=R_\infty(z)$, except, possibly near the point $Q_0$.

%%%%%%%%%%%%%%%%%%%%%%%%%%%%%%%%%%%%%%%%%%%%%%%%%%%%%%%%%%%%%%%%
\begin{remark} \lb{r4.4}
In the finite genus case, the analog of Theorem~\ref{t4.3}
was first derived by purely  algebro-geometric
means by Ehlers and Kn\"orrer \cite{EK82} in 1982.  An
elementary but rather lengthy derivation of the finite genus case
 was provided by  Ohmiya \cite{Oh97} in 1997
(based on two other papers
\cite{Oh95}, \cite{OM93}).  The present proof of
Theorem~\ref{t4.3} is patterned after the finite genus
treatment in \cite{GH98a}, which seems to be the only elementary
and short one available at this point.
\end{remark}
%%%%%%%%%%%%%%%%%%%%%%%%%%%%%%%%%%%%%%%%%%%%%%%%%%%%%%%%%%%%%%%%
\vspace*{1mm}

%%%%%%%%%%%%%%%%%%%%%%%%%%%%%%%%%%%%%%%%%%%%%
\noindent {\bf Acknowledgments.}
%%%%%%%%%%%%%%%%%%%%%%%%%%%%%%%%%%%%%%%%%%%%%
I am indebted to Helge Holden and Rudi Weikard for various
discussions on the material presented in this paper.
%%%%%%%%%%%%%%%%%%%%%%%%%%%%%%%%%%%%%%%%%%%%%%%%%%%%%%%%%%%%%%%%
%%%%%%%%%%%%%%%%%%%%%%%%%%%%%%%%%%%%%%%%%%%%%%%%%%%%%%%%%%%%%%%%

\end{document}